\documentclass[a4paper,fleqn,usenatbib]{mnras}

\usepackage{newtxtext,newtxmath}

\usepackage[T1]{fontenc}
\usepackage{ae,aecompl}

\usepackage{graphicx}	
\usepackage{amsmath}	
\usepackage{amssymb}	

\title[SPIRE Dropouts]{A SCUBA-2 Selected Herschel-SPIRE Dropout and the Nature of this Population}

\author[J. Greenslade et al.]{J. Greenslade$^{1}$ \thanks{E-mail: j.greenslade14@ic.ac.uk}, 
E. Aguilar$^{2}$, 
D. L. Clements$^{1}$,
H. Dannerbauer$^{3,4}$, 
T. Cheng$^{1}$,  \newauthor
G. Petitpas$^{5}$, 
C. Yang$^{6}$,
H. Messias$^{6,7}$, 
I. Oteo$^{8, 9}$, 
D. Farrah$^{10, 11}$,  \newauthor 
M.~J.~Micha{\l}owski$^{12}$, 
I. P{\'e}rez Fournon$^{3, 4}$, 
I. Aretxaga$^{2}$, 
M. S. Yun$^{13}$, 
S. Eales$^{14}$,  \newauthor
L. Dunne$^{8, 14}$,
A. Cooray$^{15}$ 
P. Andreani$^{9}$, 
D. H. Hughes$^{2}$. 
M. Vel\'azquez$^{2}$,  \newauthor
D. S\'anchez-Arg\"uelles$^{2}$, 
N. Ponthieu$^{16}$. 
\\
$^{1}$ Astrophysics Group, Imperial College London, Blackett Laboratory, London SW7 2AZ\\ 
$^{2}$ Instituto Nacional de Astrof\'isica, \'Optica y Electr\'onica (INAOE), Luis Enrique Erro 1, Sta. Ma. Tonantzintla, 72840 Puebla, M\'exico \\ 
$^{3}$ Instituto de Astrof{\'i}sica de Canarias (IAC), E-38205 La Laguna, Tenerife, Spain  \\ 
$^{4}$ Universidad de La Laguna, Dpto. Astrof{\'i}sica, E-38206 La Laguna, Tenerife, Spain \\ 
$^{5}$ Harvard-Smithsonian Center for Astrophysics, 60 Garden Street, Cambridge, MA 02138 \\ 
$^{6}$ European Southern Observatory, Alonso de C\'ordova 3107, Casilla 19001, Vitacura, Santiago, Chile \\ 
$^{7}$ Joint ALMA Observatory, Alonso de Córdova 3107, Vitacura 763-0355, Santiago, Chile \\ 
$^{8}$ Institute for Astronomy, University of Edinburgh, Royal Observatory, Blackford Hill, Edinburgh EH9 3HJ \\ 
$^{9}$ European Southern Observatory, Karl-Schwarzschild-Strasse 2, D-85748 Garching, Germany \\ 
$^{10}$ Department of Physics and Astronomy, University of Hawaii, 2505 Correa Road, Honolulu, HI 96822, USA \\ 
$^{11}$ Institute for Astronomy, 2680 Woodlawn Drive, University of Hawaii, Honolulu, HI 96822, USA \\
$^{12}$ Astronomical Observatory Institute, Faculty of Physics, Adam Mickiewicz University, ul.~S{\l}oneczna 36, 60-286 Pozna{\'n}, Poland \\ 
$^{13}$ Department of Astronomy, University of Massachusetts, Amherst, MA 01003, USA \\ 
$^{14}$ School of Physics and Astronomy, Cardiff University, The Parade, Cardiff CF24 3AA, UK \\ 
$^{15}$ Department of Physics and Astronomy, University of California, Irvine, CA 92697, USA \\ 
$^{16}$ Institut de Plan{\'e}tologie et d'Astrophysique de Grenoble, 414 Rue de la Piscine, 38400 Saint-Martin-d'H{\`e}res, France \\ 
} 

\date{Accepted XXX. Received YYY; in original form ZZZ}

\pubyear{2019}

\begin{document}
\label{firstpage}
\pagerange{\pageref{firstpage}--\pageref{lastpage}}
\maketitle

\begin{abstract}

Dusty star-forming galaxies (DSFGs) detected at $z > 4$ provide important examples of the first generations of massive galaxies.
However, few examples  with spectroscopic confirmation are currently known, with Hershel struggling to detect significant numbers of $z > 6$ DSFGs.
NGP6\_D1 is a bright 850 $\mu m$ source (12.3 $\pm$ 2.5 mJy) with no counterparts at shorter wavelengths (a SPIRE dropout).
Interferometric observations confirm it is a single source, with no evidence for any optical or NIR emission, or nearby likely foreground lensing sources.
No $>3\sigma$ detected lines are seen in both LMT RSR and IRAM 30m EMIR spectra of NGP6\_D1 across 32 $GHz$ of bandwidth despite reaching detection limits of $\sim 1 mJy/500 km~s^{-1}$, so the redshift remains unknown.
Template fitting suggests that NGP6\_D1 is most likely between $z = 5.8$ and 8.3.
SED analysis finds that NGP6\_D1 is a ULIRG, with a dust mass $\sim 10^8$ - $10^9$ $M_{\odot}$ and a SFR of $\sim$ 500 $M_{\odot}~yr^{-1}$.
We place upper limits on the gas mass of NGP6\_D1 of $M_{H2}$ $ < (1.1~\pm~3.5) \times 10^{11}$ $M_{\odot}$, consistent with a gas-to-dust ratio of $\sim$ 100 - 1000. We discuss the nature of NGP6\_D1 in the context of the broader submm population, and
find that comparable SPIRE dropouts account for $\sim$ 20\% of all SCUBA-2 detected sources, but with a similar flux density distribution to the general population.



\end{abstract}




\section{Introduction}

The high-redshift ($z \geq 4$) populaton of dusty star forming galaxies (DSFGs) remains poorly constrained.
Models have consistently been unable to reproduce the observed number counts of the red, $z \geq 4$ DSFGs \citep{Dowell2014, Asboth2016, Ivison2016}, and questions remain about whether DSFGs significantly contribute to the global star formation rate (SFR) density at z $>$ 3 \citep{Rowan-Robinson2016, Liu2017, Novak2017} or not \citep{Michalowski2017, Koprowski2017}. We can neither rule out a \textit{negligible} or \textit{dominant} contribution to this SFR-density from DSFGs at $z > 3.5$ \citep{Casey2018}.
Whether these mismatches are due to observational issues such as blending (eg. \cite{Scudder2016}) or lensing
 or are due to the assumptions that have gone into the numerical models (eg. \cite{Bethermin2017})
 remains unclear.
Since we expect these sources to evolve into present day elliptical galaxies in massive clusters \citep{Wilkinson2016}, the highest redshift DSFGs likely also trace the most massive dark matter halos in the early Universe. The statistical characterisation of this population is therefore a key goal for observational astronomy.

Partially, the lack of constraints on high-z DSFGs comes down to the difficulty of detecting them.
Only a handful of DSFGs at z $\geq$ 4 have spectroscopic confirmation \citep[][]{Capak2008, Coppin2009, Daddi2009, Riechers2010, Cox2011, Capak2011, Combes2012, Walter2012, Riechers2013, Dowell2014, Yun2015, Ivison2016, Oteo2017b, Oteo2016, Asboth2016, Riechers2017, Zavala2017, Strandet2017, Fudamoto2017, Marrone2017}.
The selection criteria used for these high-z DSFGs is varied; some are selected through FIR colours \citep{Riechers2013, Dowell2014, Asboth2016, Ivison2016, Zavala2017}, whilst others use mm selection \citep{Strandet2017}, and still others sub-mm wavelengths \citep{Walter2012}.
Even after candidate selection, spectroscopic confirmation remains difficult, often requiring counter-part identification in optical, near/mid infrared or radio bands, which do not benefit from the extremely negative k-correction that applies to the sub-mm and mm bands.

Most literature redshift distributions of far-infrared (FIR) and sub-mm selected DSFGs find a median redshift of $z \sim 2$, with typical interquartile ranges of $z = $ $1.8$ - $2.8$  \citep{Chapman2005, Wardlow2011, Simpson2017, Smith2017, Bakx2018}. 
Consistently, around 30\% of DSFGs have no optical/NIR/MIR/radio counterparts in these literature studies, and are routinely assumed to be high-redshift.
Surveys in the mm, such as the South Pole Telescope \citep[SPT,][]{Carlstrom2011} or AZTEC surveys \citep{Chapin2009, Vieira2013} (see also \citet{Miettinen2015}), support this, finding a median redshift distribution between 2.6 - 3.1, suggestive of a population of high-z sources, whose specific redshifts are difficult to confirm or constrain.

Arguably the most successful selection technique for high-z DSFGs has been the selection of Herschel-SPIRE 500 $\mu m$ riser sources ($S_{500} > S_{350} > S_{250}$), hereafter referred to as 500$\mu m$ risers.
This selection has led to spectroscopic confirmation of numerous $z > 4$ DSFGs \citep{Dowell2014, Asboth2016, Ivison2016}, and two $z > 6$ sources \citep{Riechers2013, Zavala2017}.
Whilst impressive, it should be noted that these results are taken from over 1000 $deg^2$ of Herschel-SPIRE data, and effectively set lower limits for the number counts of Herschel-SPIRE detectable $z > 6$ DSFGs of $ \gtrsim 2 \times 10^{-3}$ $deg^{-2}$.
Furthermore, given the confusion limited 3$\sigma$ SPIRE detection threshold of around 20 - 30 $mJy$, a source must still be, in general, highly luminous ($\sim ~ 10^{13}$ $L_{\odot}$) to be detected by SPIRE at $z > 4$, a problem which gets worse at higher redshifts.
Indeed, both of the two SPIRE detected $z > 6$ DSFGs \citep{Riechers2013, Zavala2017} are observed to have FIR luminosities $> 10^{13}$ $L_{\odot}$ (though both are additionally lensed, and only HFLS3 \citep{Riechers2013} has a FIR luminosity intrinsically $> 10^{13}$ $L_{\odot}$).

Until recently, limited field sizes and depths at sub-mm and mm wavelengths have meant that systematic searches for the rarer $z > 4$ DSFGs / sub-millimeter galaxies (SMGs) have been restricted to 500 $\mu m$ risers, lensed sources, or serendipitous discovery.
Now, however, with larger $\sim 1$ $deg^{2}$ surveys at 850 $\mu m$ such as the S2-CLS and S2-COSMOS surveys \citep[][Simpson et al. in preperation]{Geach2017}, which overlap with the larger Herschel-SPIRE extragalactic fields, new colour selections can be made.
Perhaps the most obvious high-z selection is to extend the 500 $\mu m$ risers to 850 $\mu m$ risers ($S_{850} > S_{500}$).
At a minimum, this would require detection at both 850 $\mu m$ and at least in the 500 $\mu m$ band of SPIRE to ensure the riser condition is fulfilled.
To date, only one spectroscopically confirmed 850 $\mu m$ riser is known, ADFS-27 at $z = 5.655$ \citep{Riechers2017}, suggesting this selection is reasonably successful at selecting the highest redshift DSFGs.
However, ADFS-27 is only just detected at 500 $\mu m$ in SPIRE\footnote{When including a constant confusion noise of 7 $mJy$, typical of the SPIRE maps \citep{Dowell2014, Asboth2016}}, with a flux density of $S_{500} = 24.0 \pm 7.5$, and is not detected in either of the other two SPIRE bands.

The apparent rarity of 850 $\mu$m risers detected at both 500 $\mu m$ and 850 $\mu m$ makes systematic selection of high-z DSFGs difficult.
However, catalogues of sources from these new, $> 1$ $deg^2$ fields at 850 $\mu m$ have revealed a population of reasonably bright ($> 5 mJy$) 850 $\mu m$ sources that are not only \textit{not} detected at other optical, near/mid-infrared or radio wavelengths, but are additionally undetected in any of the three Herschel-SPIRE bands.
These ``SPIRE dropouts'' are difficult to explain, as, unless the peak of the thermal emission lies near 500 $\mu m$ - 850 $\mu m$, we would expect to detect them in the shorter wavelength SPIRE bands. 
The nature of these dropouts is uncertain, but the two simplest explanations are that this population is either very high redshift, with a median redshift higher than the 500 $\mu m$ risers, or that they represent a cooler population hitherto undiscovered at $z > 4$.
Both of these solutions are interesting in their own right, indicating that these dropouts are worthy of further study.

In this paper, we detail NGP6\_D1, a serendipitous SPIRE dropout first identified in 2014, our subsequent follow up, and our interpretation of what this source, and others like it, represent.
Given the numerous observations taken of NGP6\_D1, we start in Section 2 by providing an overview of all the observations that have taken place. We then present the photometric and spectroscopic analysis for this source in Section 3.
In Section 4, we discuss NGP\_D1 in the broader context of the current searches for high-z DSFGs, including comparisons to the literature, and examine how numerous NGP6\_D1-like objects might be, and what they might represent.
Finally, we summarise and conclude our results in Section 5.
Throughout this paper, we assume the concordance $\Lambda$-CDM cosmology, with H$_0 = 67.74$ km s$^{-1}$ $Mpc^{-1}$, $\Omega_{\Lambda} = 0.69$ and $\Omega_{m} = 0.31$.

\section{Data}
\label{sec:data}


NGP6\_D1 was initially detected serendipitously, as part of a follow up program of Planck selected Herschel overdensities \citep[][Cheng et al. in preperation]{Clements2014, Greenslade2018}.
A region in the North Galactic Pole (NGP), initially observed by Herschel as part of the H-ATLAS project \citep[][]{Eales2010, Bourne2016b, Valiante2016, Smith2017}, was observed by the SCUBA-2 instrument on the JCMT at 850 $\mu m$ (Project ID: M13AU12) between April 8th and 12th 2013.
The observations used a CV\_DAISY pattern, and reached an approximately uniform rms of $\sim3~mJy$ over a 2 arcminute radius.
The atmospheric opacity on the nights varied between $\tau_{225~GHz} = 0.05 - 0.12$, and pointing was done using the quasar 1308$+$326. 
The data were reduced using the SCUBA-2 pipeline \textsc{SMURF} \citep{Chapin2013}, and for calibration the standard 850 $\mu$m Flux Conversion Factor (FCF) of 537 $Jy~ pW^{-1}$ was used.

\begin{figure}
    \centering
    \includegraphics[width=1\linewidth]{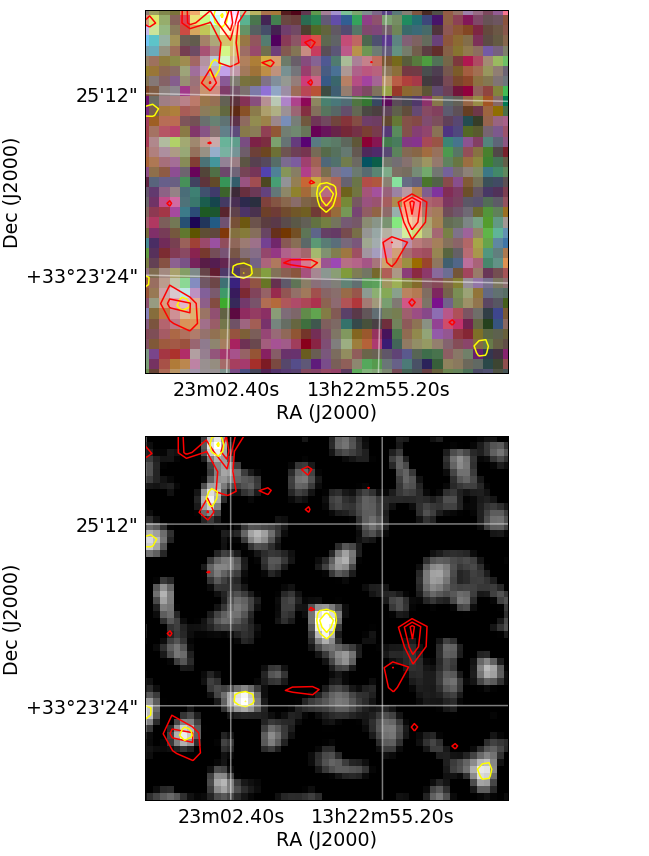}
    \caption{\textit{(Top)} A 2'.5 $\times$ 2'.5 RGB (500, 350 and 250 $\mu m$) Herschel-SPIRE image of the region around NGP6\_D1. Red contours show the 3, 4 and 5$\sigma$ detection levels for the 500 $\mu m$ band, whilst yellow contours show the 3, 4 and 5$\sigma$ detection levels for the SCUBA-2 850 $\mu m$ band. \textit{(Bottom)} The SCUBA-2 image of the same region, with contours indicating the same as the top image.}
    \label{fig:SCUBA2AndHerschel}
\end{figure}

This map revealed a 12.3 $\pm$ 2.5 $mJy$ SCUBA-2 source detected at 4.9$\sigma$ at position RA : 13:22:57.91, Dec : +33:24:14.05.
After examining the Herschel-SPIRE maps at the position of this SCUBA-2 source, we found no evidence of any emission in any of the three SPIRE bands, with measured flux densities of -2.31 $\pm$ $5.54$, 2.30 $\pm$ $5.84$ and 7.49 $\pm$ $7.35~mJy$ at 250, 350 and 500 $\mu m$ respectively.
Given a 3$\sigma$ detection limit, this places upper limits on the Herschel-SPIRE flux density of NGP6\_D1 of 16.5, 17.4, and 22.0 mJy at 250, 350, and 500 $\mu~m$.
Bootstrapping the SCUBA-2 data revealed that the source was likely real, and still detected to at least a 3$\sigma$ level when randomly discarding half the data. 
The nearest detected \textit{Herschel} source lies 45 arcsec away from the peak of the SCUBA-2 emission.
A Herschel RGB (500, 350, 250 $\mu$m) map with the contours overlaying the SCUBA-2 position is shown in Figure \ref{fig:SCUBA2AndHerschel}.
As this source has effectively dropped out of the SPIRE bands, we herein refer to it as a SPIRE dropout, with the designation NGP6\_D1.

In this Section, we detail our photometric and spectroscopic follow up observations of this isource, with a summary of our observations available in Table \ref{table:photometry}.

\begin{table}
    \caption{The FIR, sub-mm, mm and radio photometry of NGP6\_D1. Flux densities are given in $mJy$. These measurements come from Herschel-SPIRE (250, 350, 500 $\mu m$), SMA (870 $\mu m$, 1.1 $mm$ ), NIKA (1.25 and 2.0 $mm$) NOEMA (2.8 $mm$), and VLA (6 $GHz$) at the position of the source as derived from the SMA maps. Square brackets indicate 3$\sigma$ upper limits in the case of non-detection in the SPIRE maps; specific SPIRE values at the position of the source are included as these are used for template fitting to constrain the peak of the thermal emission.}

    \centering
    \begin{tabular}{c|c}
        \hline
        \hline
         Band & Flux [$mJy$] \\
        \hline 
         250 $\mu m$ & -3.3 $\pm$ 4.2  $[< 12.6]$\\
         350 $\mu m$ & 3.0 $\pm$ 4.4 $[< 13.2]$\\
         500 $\mu m$ & 7.7 $\pm$ 8.9 $[< 26.7]$\\
         850 $\mu m$ & 12.3 $\pm$ 2.5 \\
         870 $\mu m$ & 8.0 $\pm$ 1.3 \\
         1.1 $mm$ & 5.9 $\pm$ 1.1 \\
         1.25 $mm$ & 3.97 $\pm$ 0.43 \\
         2 $mm$ &  1.04 $\pm$ 0.12 \\
         2.8 $mm$ & 0.60 $\pm$ 0.04 \\
         6 $GHz$ & (1.69 $\pm$ $0.4)  \times  10^{-2}$ \\

         \hline
         
    \end{tabular}
    \label{table:photometry}

\end{table}

\subsection{Photometric Observations}

\subsubsection{SMA}

We undertook observations using the Sub-Millimeter Array (SMA) in extended configuration, at 870 $\mu$m and 1.1 mm on March 29$^{\textit{th}}$ 2015 and March 23$^{\textit{rd}}$ 2015 respectively (Project ID: 2014A-S092). 
The bandpass calibrator was 3c84, while Callisto was used as flux density calibrator, and the quasars 1310+323, and 1224+213 were used as gain calibrators.
The data was reduced with a combination of both \textsc{IDL} and \textsc{MIRIAD} using natural weighting to optimise the point-source sensitivity.
The smaller 870 $\mu m$ synthesised beam had semi-major and semi-minor axis of 0''.78 $\times$ 0''.47 and the maps reached 1$\sigma$ rms of 1.31 and 1.36 $mJy$ in the 870 $\mu m$ and 1.1 $mm$ bands respectively.

\begin{figure}
    \centering
    \includegraphics[width=1\linewidth]{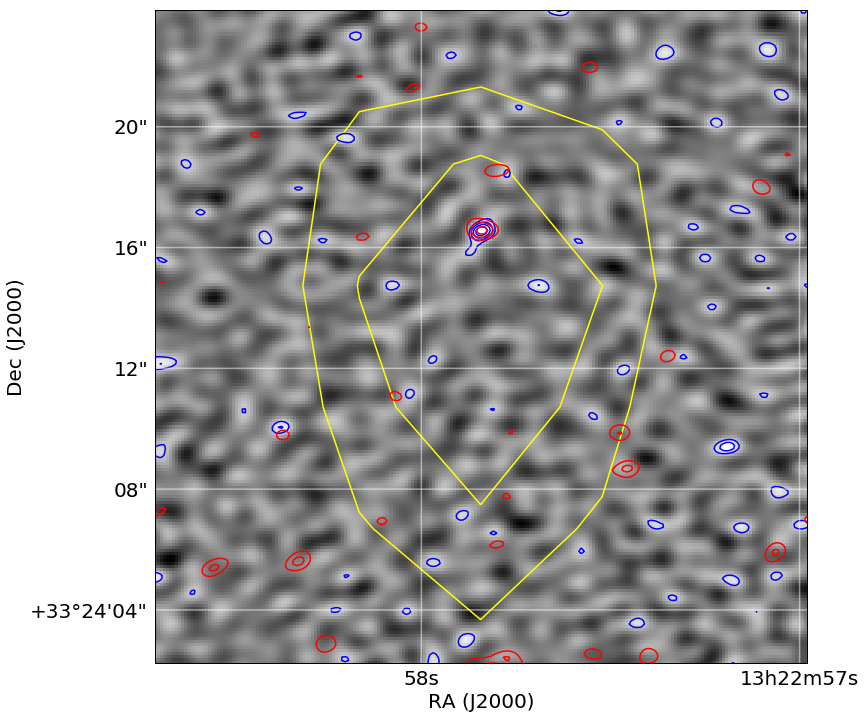}
    \caption{The SMA 345 $GHz$ (870 $\mu m$) map of NGP6\_D1, overlaid with SCUBA-2 S/N contours (yellow) in steps of 3, 4, and 5$\sigma$, and SMA 870 $\mu m$ S/N contours (blue) and SMA 1.1 $mm$ contours (blue), both in steps of 2, 3, 4, and 5$\sigma$.}
    \label{fig:SMANGP6}
\end{figure}

In both the 870 $\mu m$ and 1.1 mm maps, a source was found, well within the $\sim 13$ arcsec full-width half-maximum (FWHM) SCUBA-2 beam, and with position RA: 13:22:57.842, and Dec: +33:24:16.56.
The measured flux densities were 8.03 $\pm$ $1.31~mJy$ and 5.96 $\pm$ $1.36~mJy$ at 870 $\mu m$ and 1.1 $mm$ respectively.
The 870 $\mu m$ flux density values are consistent within 2$\sigma$ with the observed SCUBA-2 flux density.
The SMA image of NGP6\_D1 at 870 $\mu m$ is plotted in Figure \ref{fig:SMANGP6}, alongside the contours at both 1.1 $mm$ from the SMA and 850 $\mu m$ from SCUBA-2.
The Herschel measured flux densities at the position of the SMA source are -3.34 $\pm$ 4.16 $mJy$, 2.98 $\pm$ 4.42 $mJy$, and 7.70 $\pm$ 8.90 $mJy$ in the 250, 350 and 500 $\mu m$ bands respectively.

\subsubsection{NIKA}

NGP6\_D1 was observed on the IRAM 30m telescope using the NIKA \citep{Monfardini2010} instrument at $1.25$ and $2$ $mm$ (beamsizes of 12 and 17.5 arcsec) between the 8$^{th}$ and 9$^{th}$ of February 2015 (Project ID: 227-14) for 2 hours, reaching rms values of $\sim$ 0.4 $mJy$ and 0.1 $mJy$ in the two bands respectively. 
Tau values ranged between $\tau_{225~GHz} = 0.01$ and $0.28$ and with an average of 0.15, but this was generally split between a high opacity $\tau_{225} > 0.1$ and low opacity $\tau_{225} < 0.1$ grouping.
The data were reduced by the NIKA team's pipeline, using a ``point source oriented'' reduction.
A single source was found at the position of the SMA object, with a 1.25 $mm$ flux density of 3.97 $\pm$ 0.43 $mJy$ and a 2 $mm$ flux density of 1.04 $\pm$ 0.12 $mJy$.
The fluxes were found to be consistent when using only the high tau or low tau datasets, but there remains a 10 - 15\% uncertainty on the flux calibration. 
The 1.25 $mm$ flux from NIKA appears inconsistent with the 1.1 $mm$ flux from the SMA, with the SMA 1.1 $mm$ measurement 50\% higher than the NIKA 1.25 $mm$ measurement. 


\subsubsection{VLA}

The region around NGP6\_D1 was observed by the Very Large Array (VLA) on 17 December 2016 (Project ID VLA/2016-00-110, PI David Clements) for 1.75 hours.
This observation was in C-band (6 $GHz$, 50 $mm$), and in A-configuration, with a synthesised beamwidth of 0$.\!\!{''}$33 $\times$ 0$.\!\!{''}$33. 
For bandpass and flux calibration, 3C286 (1331+305) was used, while J1310+3220 was the phase calibrator. 
Data were reduced using the Common Astronomy Software Application (\textsc{CASA}) package \citep{McMullin2007}, version 4.7.0.
Small amounts of radio frequency interference (expected to be around $\sim$ 15\%) were detected and flagged automatically during the reduction process.
The field is cleaned using Briggs (robust) weighting, with a robust parameter of 0, to provide a good balance between angular resolution and sensitivity to all sources in the field. 
These radio observations were taken on a number of protocluster candidates, and the full results will be presented in a future paper (Cheng et al. in preparation).

We examined the map around the position of NGP6\_D1 and detected a 4.5$\sigma$ source, with a 6 $GHz$ flux density of 16.9 $\pm$ $3.7$ $\mu~Jy$, with a position only 0.05'' from the SMA position of NGP6\_D1.
Assuming the radio emission is concurrent with the FIR emission, our VLA map localises our source to a 0''.33 $\times$ 0''.33 area on the sky.

\subsubsection{Ancillary data from SDSS, UKIDSS and WISE}

The area around NGP6\_D1 was observed in both the Sloan Digital Sky Survey  (SDSS, \citep{Abolfathi2017}) and UKIRT Infrared Deep Sky Survey (UKIDSS, \citep{Lawrence2006, Warren2006}) in the optical and near-infrared (NIR).
These observations reached approximate AB magnitude limits of $u:22.0$, $g:22.2$, $r:22.2$, $i:21.3$ and $z:20.5$ in SDSS, and Vega limits of $Y:20.2$, $J:19.6$, $H:18.8$, and $Ks:18.2$ from UKIDSS.
Though there are two SDSS galaxies approximately 8 and 11 arcsec to the north of the SMA positions of NGP6\_D1, there is no current evidence of any optical counterpart, or indication that NGP6\_D1 is being lensed by any foreground source. 
We do note however that deeper images in the optical / NIR may change this.

\subsection{Spectroscopic Observations}

Through both photometric analysis and template fitting of the above data (see Section \ref{sec:results} for more details), we estimate that the most likely redshift for NGP6\_D1 is between $z = 5.8 $ - $8.3$.
To determine a redshift, we opted to target $^{12}$CO lines (hereafter referenced as simply CO).
The CO(J = 1 $\rightarrow$ 0) transmission occurs at a rest-frame frequency of 115.27 $GHz$ ($\sim$ 2600 $\mu m$), with subsequent CO(J = n $\rightarrow$ (n-1)) transmissions taking place at n $\times$ 115.27 $GHz$.
At $z > 5$, we would therefore expect adjacent CO lines to be spaced out by $ < 20$ $GHz$.
Blind redshift searches on SMGs targetting CO lines have been performed before, and to a reasonable level of success \citep{Weiss2013, Dannerbauer2014, Fudamoto2017}.
In this Sub-section, we report our spectroscopic observations of NGP6\_D1 using both the RSR and EMIR instruments, and our resulting spectra from both instruments are shown in Figure \ref{fig:RSREMIR}.

\subsubsection{Redshift Search Receiver} \label{sec:rsr}

The Redshift Search Receiver (RSR, \citealt{erickson07}, \citealt{goeller08}) is the wide-band 3 mm spectrograph currently installed on the 50-m Large Millimeter Telescope (LMT, \citealt{hughes10}). 
It has a spectral resolution of 31 $MHz$ or $\sim$ 100 $km/s$ at 92 $GHz$, and an instantaneous frequency coverage of 73 to 111 $GHz$. 
The RSR follow-up of NGP6$\_$D1 was conducted on the Early Science phase with a 32-m dish configuration which provides a spatial resolution of 25 $arcsec$ at 92 $GHz$. 
The opacity $\tau_{225GHz}$ ranged between 0.10 and 0.27 with an average T$_{sys} \sim$ 100 $K$ over the 6 observation nights (2016 January 29 and February 1-3, 7 and 8).  
The total on-source integration time on NGP6\_D1 was 9 hours (108 spectra $\times$ 300 seconds each). 
Pointing corrections were made observing 1224+213 or 1310+323 every hour. 

The individual observations are transformed into the frequency domain, baselined and co-added using DREAMPY (Data REduction and Analysis Methods in PYthon2), written by Gopal Narayanan, to generate the spectrum.
The final spectrum was obtained by co-adding the best data, defined as all the individual spectra which do not have large structure systematics in the baseline due to low frequency noise (electronic drift).
After co-addition the data are smoothed with a 3 channel boxcar filter. 
Additionally, we smooth the coadded spectrum to match a velocity resolution of 500 km s$^{-1}$, typical of other high-z DSFGs \citep{Riechers2013, Bothwell2013, Aravena2016, Zavala2017, Strandet2017, Yang2017}. 
To convert from antenna temperature to flux units we use a factor of 6.4 $Jy/K$ for $\nu \leq$ 92 $GHz$ and 7.6 $Jy/K$ for $\nu$ $>$ 92 $GHz$.

\begin{figure*}
    \centering
    \includegraphics[width=1\linewidth]{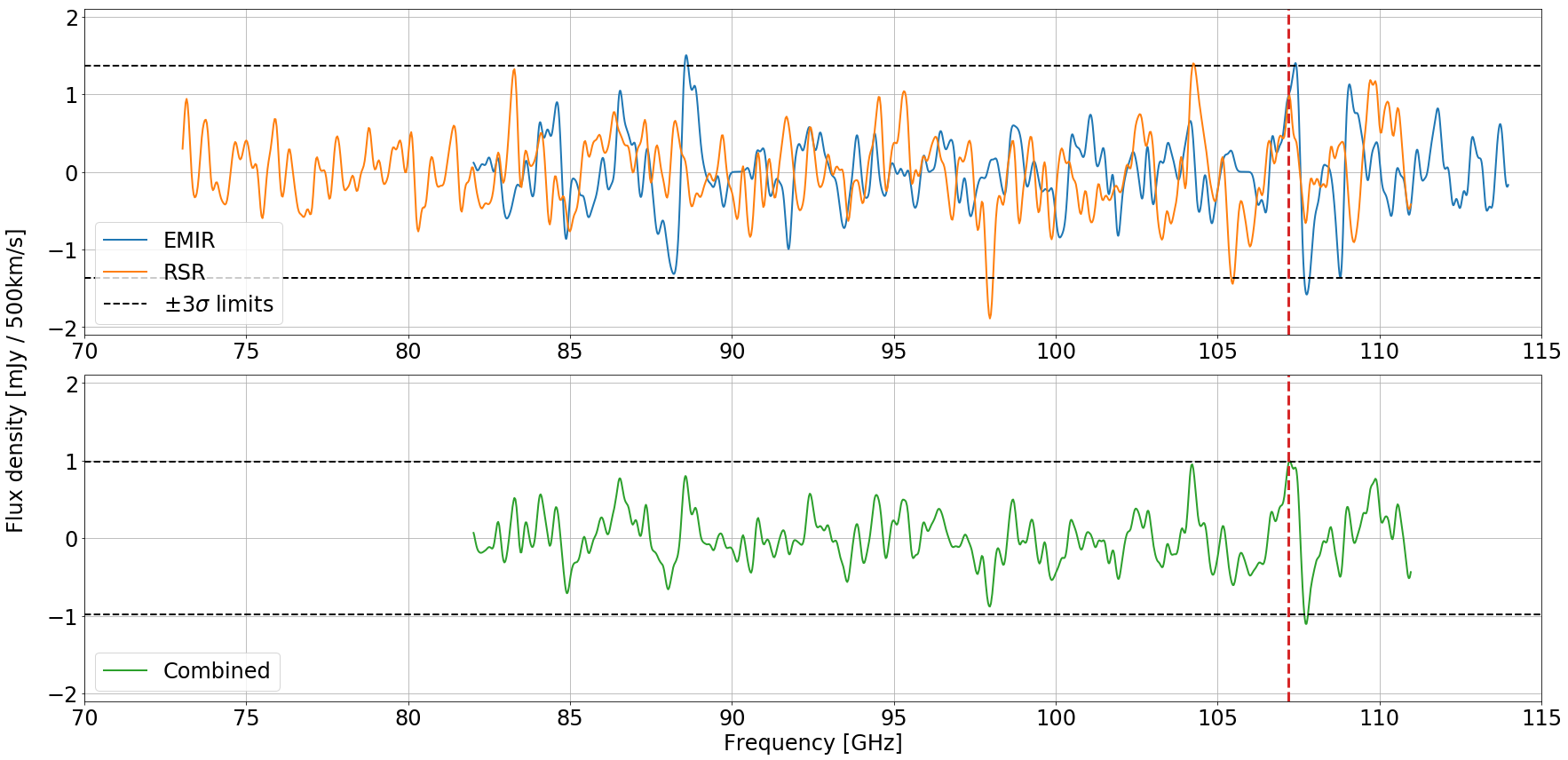}
    \caption{Spectra of NGP6\_D1 from the RSR and EMIR (top), and the spectra from combining the data from the two instruments (bottom). The data is smoothed with a Gaussian with a velocity width of 500 km/s in all three cases. The dashed horizontal black lines show the $\pm$3$\sigma$ limits for both the RSR, EMIR and combined, whilst the dashed red vertical line shows the position of a 3.03$\sigma$ line at 107.2 GHz when the data from the RSR and EMIR are combined and smoothed to a velocity width of $\sim$ 500 km/s. }
    \label{fig:RSREMIR}
\end{figure*}

As can be seen in Figure \ref{fig:RSREMIR}, there is a $>3\sigma$ feature detected at 104.28 $GHz$ in the RSR spectrum, and a second line marginally detected to a 2.8$\sigma$ level at 83.2 $GHz$. 
This could reasonably correspond to a $z \sim 4.53$ SMG, and template cross-correlation analysis (i.e. \citet{Yun2015}) suggests a combined detection S/N of 5.5 in support of this redshift solution. 
However, as discussed below, neither candidate line is detected in EMIR, and there are additionally two frequencies that are negatively detected as strongly as these candidate lines.
Further evidence would be needed before any definite conclusion as to the reality of these lines can be made, and we therefore conclude that there is no strong evidence for any detected features in the RSR spectrum.




\subsubsection{EMIR}
NGP6\_D1 was observed for a total of 61 hours with the EMIR instrument on the IRAM 30m telescope (Project ID: 199-15) to search for $^{12}$CO lines.
Two setups covered a total of 31 GHz (83 - 114GHz) of frequency space to an rms of 0.06 mK ($\sim 0.42$ mJy), with two small 1GHz gaps at 90 and 105.5 GHz due to different set ups.
The observations ran from the 14$^{th}$ of March to the 20$^{th}$ of March 2016, with tau values varying from 0.01 to 0.5, with an average of $\tau_{225}\sim 0.2$.
Both WILMA and FTS200 were used as back-ends, with FTS200 covering a larger 32 GHz of bandwidth compared to WILMA.
The data were reduced using CLASS and Python, and smoothing our data to between 100 and 500 km/s, we achieved an rms of between 0.1 and 0.07 mK, corresponding to a line sensitivity of 0.7 - 0.5 mJy.
No evidence of any lines is seen in the WILMA back-end, but FTS200 covering a larger bandwidth detect two $\sim 3\sigma$ peaks, one of which appears concurrent with a peak in the RSR spectra.

\subsubsection{NOEMA}

Both EMIR and the RSR see a marginal line at 107.2 GHz, with $\sim$3$\sigma$ and 1.7$\sigma$ detections respectively.
Combining these results together, as seen in Figure \ref{fig:RSREMIR}, results in a 3.2$\sigma$ detection of this line.
We therefore obtained NOEMA DDT to follow up this candidate line.

\begin{figure}
    \centering
    \includegraphics[width=1\linewidth]{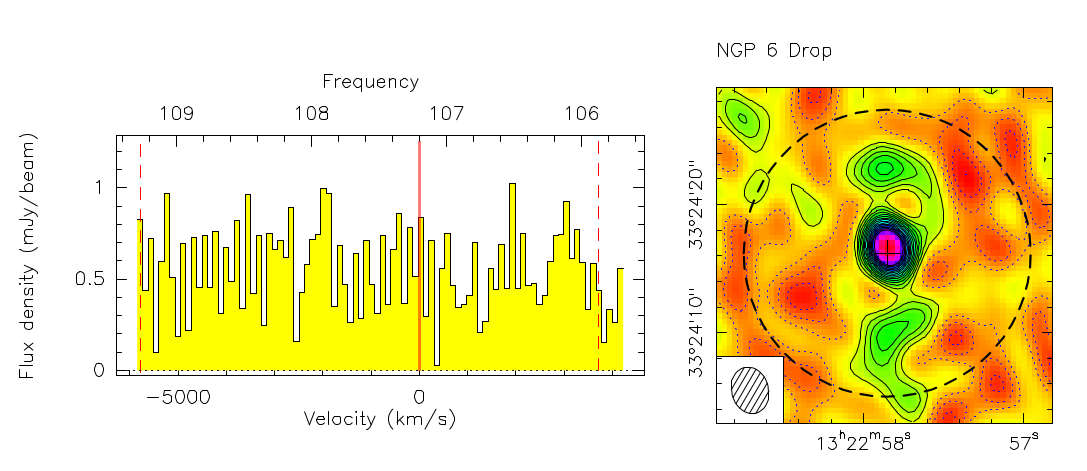}
    \caption{ (Left) The NOEMA spectra surrounding the candidate line at 107.2 GHz. The solid red vertical line indicates the position of the marginal line seen in Figure \ref{fig:RSREMIR}. (Right) The dirty map of NGP6\_D1. The northern and southern side-lobes seen are artefacts from the beamshape of NOEMA, and do not represent emission. Contours are in steps of 1$\sigma = 0.24 $ Jy km/s beam$^{-1}$}
    \label{fig:NOEMA}
\end{figure}

The NOEMA Interferometer is a millimeter array located on the Plateau de Bure in the French Alps.
A spectral line scan of NGP6\_D1 was carried out in January 2017 (DDT E16AD: PI J. Greenslade) with 7 (20 January 2017) and 8 antennae (21 January 2017) in D configuration to search for the possible line at 107.2 GHz.
The Wide-X receiver was used, which provides a bandwidth of 3.6~GHz. The data were calibrated through observations of standard bandpass (3C84, 1055$+$018), phase/amplitude (1328$+$307, J1310$+$323) and flux density calibrators (LKHA101, MWC349) and reduced with the GILDAS software packages CLIC and MAP. The FWHM of the beam was $3.8''\times3.0''$ at 107.2~GHz, slightly larger than the SMA beamsize.
The continuum and spectrum is shown in Figure \ref{fig:NOEMA}, with the red line indicating the expected position of the line.

No line was found at 107.2 GHz, indicating the candidate line was just a noise spike, and highlighting the difficulty in obtaining spectroscopic redshift confirmations of these faint sub-mm sources.
However, we did detect the continuum emission of NGP6\_D1, with a flux at $S_{107.2~GHz} of 0.56$ $\pm$ $0.03$~mJy, an 18.6$\sigma$ detection. 
The derived position is at RA$=$ 13:22:57.837 DEC$=+$33:24:16.61 (J2000), only $0.4''$ away from the pointing centre (SCUBA-2 position), and only 0.05'' from the SMA position. 

\section{Results}

\label{sec:results}



         

\begin{figure}
    \centering
    \includegraphics[width=1\linewidth]{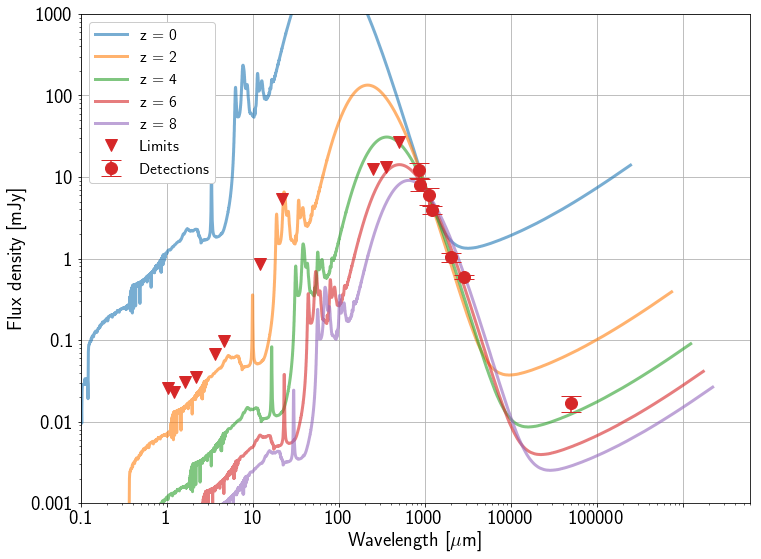}
    \caption{The ALESS average SED, as it would appear at $z = 0,$ 2, 4, 6, and 8 if normalised each time to the observed 1.1 $mm$ flux density of NGP6\_D1. The red triangles shows photometric bands where we only have upper limits, whilst the points with errorbars show where we have $>$3$\sigma$ detections.}
    \label{fig:aless}
\end{figure}

NGP6\_D1 is detected in sub-mm, mm, and radio photometric bands, but no optical, NIR or FIR bands.
We have further determined it is likely either a single source, or very close merger, with on-sky separation of $< 2$''.
Despite this, we have been unable to determine the redshift of NGP6\_D1.
Well-studied local ULIRGs and SMGs are often used as templates when fitting photometric redshifts, under the assumption that the template SED is well matched to the underlying SED of the source \citep{Ivison2016, Ikarashi2017, Duivenvoorden2018}.
Under this assumption, in Figure \ref{fig:aless}, we plot a representative SMG SED (the ALESS average SED, \citealt{daCunha2015}) at redshifts of 0, 2, 4, 6 and 8, normalising each time to the 1.1 $mm$ detection (arbitrarily) of NGP6\_D1.
We then over-plot our optical/NIR limits, and observed sub-mm, mm and radio detections.
As can be seen, the lack of a SPIRE detection immediately implies very red sub-mm colours for NGP6\_D1; at any redshift below $z = 4 - 6$, we would expect to detect NGP6\_D1 in at least one of the SPIRE bands, and at $z < 2$ likely in the optical and NIR bands as well.
The SPIRE photometry is consistent with NGP6\_D1 being at least a 500 $\mu$m riser, if not a 850 $\mu$m riser.
In this Section, we attempt to estimate a photometric redshift for NGP6\_D1, and use this to derive a luminosity.
Furthermore, we will examine the radio detection and CO limits, and their implications for the dust mass of NGP6\_D1.

\subsection{Redshift estimates}

Constraining the redshift of individual FIR / sub-mm detected objects is notoriously difficult.
The two most common options include fitting single modified blackbodies with and without priors \citep{Greve2012, Weiss2013, Riechers2013, Strandet2016}, and template fitting using a single or a library of templates \citep{Lapi2011, Pearson2013, Ivison2016}.
Both options require a questionable set of assumptions, in the case of fitting single temperature blackbodies the assumption that a single temperature fits the true SED well (see \cite{Strandet2017} for a counterexample), and in the case of template fitting that one or any of the templates are well matched to the SED of the source.
Given the SPIRE-dropouts are poorly studied in general, we can not be certain that either of these assumptions are valid here.
Furthermore, as we only have weak constraints from our Herschel non-detections on the most prominent feature of the FIR SED, the frequency peak of the SED, it is prudent to be conservative in our estimates of the redshift.
As such, we opt to use both methods, whilst adopting broadly conservative priors and template libraries, so as to correctly reflect our ignorance.

\subsubsection{Fitting templates}

\label{sec:tem}

Whilst a single template may not accurately reflect the SED of a single source, a broad range of templates that span a larger range of parameter space will likely capture the true uncertainty in the redshift of a source. This procedure was thoroughly investigated as applied to DSFGs in \citep{Ivison2016} which demonstrated its effectiveness in recovering the redshifts of DSFGs with known known spectroscopic redshifts\footnote{It is worth noting however that this assumption is not always valid, even when using numerous templates; \citet{Ikarashi2017} fit SMGs from a parent sample of 185 SED templates, and whilst able to accurately fit most of their sources, they are still unable to find a good fit for HFLS3, which they ascribe to HFLS3's warm dust temperature (Section 4.1 and 4.3 of \citet{Ikarashi2017}.}.
To estimate the redshift of NGP6\_D1, we utilise eight separate templates which host a broad range of properties: Cosmic Eyelash \citep{Ivison2010, Swinbank2010, Danielson2011}, ALESS average \citep{daCunha2015}, Arp 220 \citep{Rangwala2011}, M82, NGC 6090, IRAS 20551-4250, IRAS 22491-1808, and two sources with known AGN, Mrk 231, and a QSO template.
The last six of these are all from the \citealt{Polletta2007} library of SEDs.

\begin{figure*}
    \centering
    \includegraphics[width=1\linewidth]{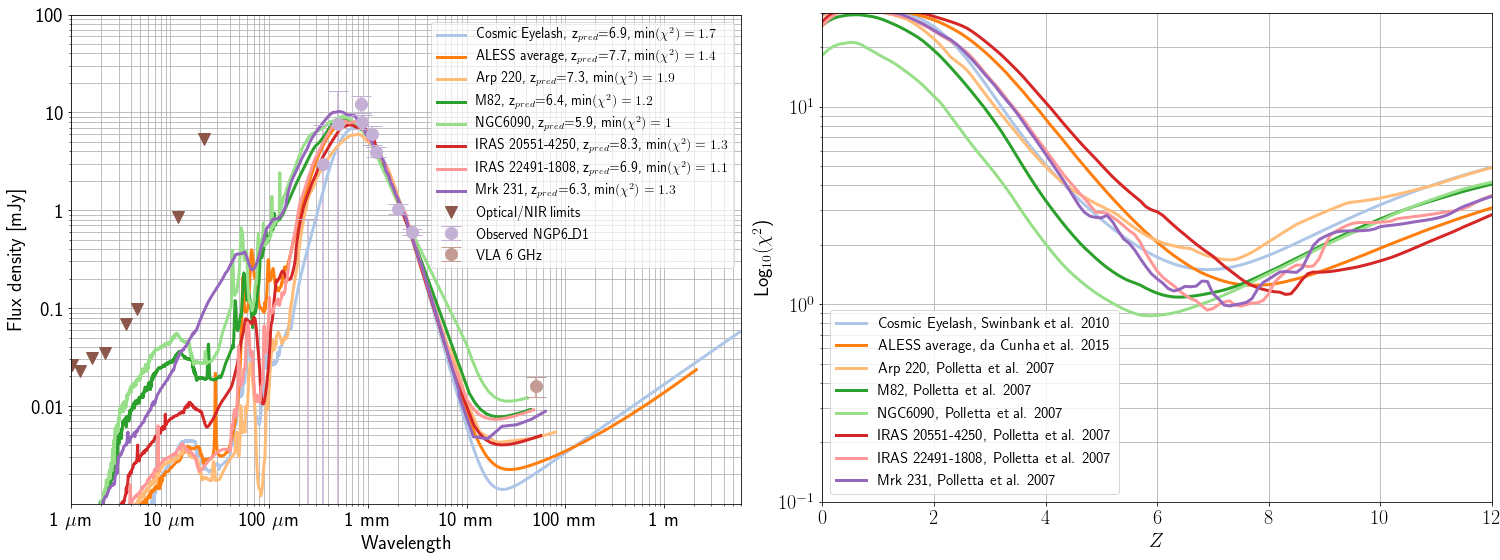}
    \caption{\textit{Left)} The best fit redshift and amplitude for eight templates fit to the photometry of NGP6\_D1. The solid coloured lines show the different templates, whilst the legend shows the best fit redshift and minimum $\chi^{2}$. The grey points with errorbars show the photometry used in the fits, the pink point shows the radio point not used in the fit, and the brown triangles show the optical / NIR upper limits. \textit{Right)} The $\chi^2$ as a function of redshift for all of the templates used in our fitting. The colours correspond to the plot on the left, and the source for each of the templates is provided in the legend.}
    \label{fig:SED}
\end{figure*}

On the left of Figure \ref{fig:SED}, we plot the best fit redshift and normalisation for each of the eight templates.
We use the photometry given in Table \ref{table:photometry}, excluding the radio point since not all templates include radio data.
In each case and for each template, we minimize the $\chi^2$ between the template and our data, allowing both the normalisation amplitude and redshift to vary.
This gives, for each template, a best fit redshift for NGP6\_D1.
We additionally plot the $\chi^2$ as a function of redshift for the procedure on the right of Figure \ref{fig:SED}, showing that each template performs similarly and additionally highlighting the reasonably broad minimum for each template.
To obtain a likely redshift range, we take both the template with the lowest best fit redshift and the template with the highest best fit redshift (in this case NGC6090 and IRAS 20551-4250 respectively), and use this range as the likely redshift range appropriate for NGP6\_D1.
We stress this is specifically \textit{not} an error range, which would slightly extend this range beyond its limits, but is a range of best fit redshifts, given a broad range of templates from the literature.

For NGP6\_D1, the best fit redshift ranges from a minimum of $z = 5.88$ to a maximum of $z =$ 8.33, with a mean and median redshift estimate from all the templates around $z \sim 6.9$.
Assuming the true redshift lies somewhere within this range, this implies that NGP6\_D1 is likely one of the highest redshift DSFGs found to date.
The reduced $\chi^2$ values range between $\chi_\nu^2 = 0.125$ - 0.24, indicating that in all cases we are generally over-fitting the models.
This is not surprising given the lack of informative features in the long wavelength tail of the dust SED; a single detection in the optical, near-infrared, or mid-infrared would significantly help constrain the true redshift.

\begin{figure}
    \centering
    \includegraphics[width=1\linewidth]{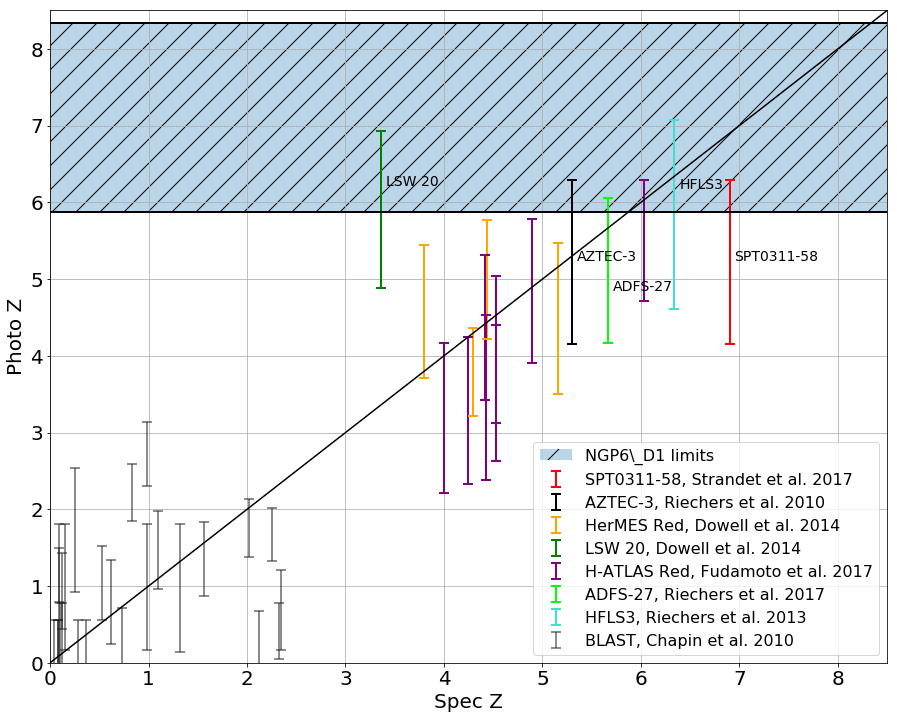}
    \caption{A comparison between the photometric and spectroscopic redshift of a variety of literature DSFGs, with photometric redshift fit using the method described in the text. Origins of the data are shown in the legend, whilst several well known DSFGs are labelled with text to their right. The blue shaded region represents the photometric redshift estimate for NGP6\_D1.}
    \label{fig:photozspecz}
\end{figure}

However, does such a fit generally contain the true redshift of the source? 
To test this, we searched for all the examples we could find of $z > 3$ DSFGs with spectroscopic confirmation and similar observed photometry to NGP6\_D1, and ran those sources through our template fitting procedure. 
We additionally selected a number of sources from the BLAST survey \citep{Chapin2011} to test that our procedure would also correctly identify lower redshift $z < 3$ sources.
In Figure \ref{fig:photozspecz} we plot the results of our fitting procedure to all similar high-z sources in the literature, as well as sources from the BLAST survey.
The data used for fitting in each case broadly matches those we have for NGP6\_D1 (i.e. the Herschel SPIRE bands plus a number of sub-mm and mm bands where available).
As can be seen, in almost all cases the true redshift is contained within the min-max range given by the fits. 
The only exceptions to this, excluding the BLAST sources which are generally only detected in 1 - 2 FIR bands, are SPT-0311-58 \citep{Strandet2017}, and LSW 20 \citep{Dowell2014}, which are under and over predicted respectively. The reasons for these discrepancies are not clear; both are 500 $\mu m$ risers, and both have dust temperatures between $\sim 40 - 60$K\footnote{It is important to note that SPT0311-58 is poorly fit by a single temperature, and indeed \citet{Strandet2017} use a two-component model, with a cold and warm dust temperature of 36 $\pm$ 7 K and 115 $\pm$ 54 K respectively.}.
These errors indicate the inherent difficulty in fitting photometric redshifts from templates, but it is encouraging that all the other $z > 3$ sources are well fit by our choice of templates.
Nevertheless, the possibility that NGP6\_D1 could be similar to LSW-20 or SPT0311-58, and possibly lie at a lower or higher redshift than predicted, cannot be discarded.

Comparing the predictions for NGP6\_D1 to the other high-z literature DSFGs, it is clear that NGP6\_D1 is predicted to lie at a higher redshift than all other known sources. 
Its low redshift estimate at $z = 5.9$ is already higher than the highest redshift estimate for all but 6 sources. 
Its high redshift estimate at $z = 8.3$ is higher than any other high redshift estimate for any other source.
As detailed above, there are many uncertainties to these fits, but in general fitting to templates favours a high-z $z > 5$ solution for NGP6\_D1.


However, if our source is much cooler than, or has an SED intrinsically different to, the templates used here, then our templates will be poor fits and are unlikely to correctly identify the redshift of the source.
LSW 20 is a good example of where this fitting procedure fails (see \citealt{Dowell2014} for more extensive examination of LSW 20), and if NGP6\_D1 is similar to LSW 20 then we may expect NGP6\_D1 to have a redshift significantly lower than predicted here.
In the next Section, we therefore look to fitting single modified blackbodies to our source, which can have a broad range of temperatures and redshifts, and examine at any given redshift, what dust temperatures our source would need to posses, and whether these are physically sensible.


\subsubsection{Fitting single modified blackbodies}

\label{sec:black}

In order to model the thermal emission from NGP6\_D1, we assume the FIR spectrum is well represented by a single dust temperature modified blackbody \citep{Blain2002, Magnelli2012, Bianchi2013, Casey2014}.
This model usually takes the form:
\begin{equation}
    \label{eq:MBB}
    S_{\nu} \propto (1 - exp(-\tau_{\nu})) B_\nu,
\end{equation}
where $S_{\nu}$ is the observed flux density at frequency $\nu$, $\tau_{\nu}$ = $(\frac{\nu}{\nu_{0}})^{\beta}$, and gives the optical depth at frequency $\nu$, $\nu_{0}$ is the frequency at which the optical depth equals unity, and $B_{\nu} = B_{\nu}(\nu, T_{dust})$ is the Planck function.
$\beta$, is usually assumed to be $\beta = 1.5$ - 2 for SMGs \citep{Blain2002, Casey2014}.
In this model, there are five parameters to be fit: The redshift $z$, the average dust temperature $T_{dust}$, the dust emissivity $\beta$, the frequency at which the optical depth reaches unity $\nu_{0}$, and an overall normalisation parameter $a$.

To fit our data to this model, and similar to \citet{Dowell2014} and \citet{Asboth2016}, we use the affine invariant Markov Chain Monte Carlo \citep{Goodman2010} ensamble sampler \textsc{Python} package, \textsc{emcee} \citep{ForemanMackey2013}.
We use the following uninformative priors for our parameters: 0 $ < z \leq $ 12, $T_{CMB}(z) \leq T_{dust} \leq 80$, $1 \leq \beta \leq 3$, 1 $\mu m$ $ \leq c/\nu_{0} \leq 1$ $mm$, and $-2 \leq log_{10}(a) \leq 2$, where $T_{CMB}$ gives the CMB temperature at redshift $z$, and $c$ gives the speed of light.
For numerical stability, at each sample we first normalise to the 850 $\mu m$ observation, and allow the normalisation $a$ to vary from there.
Tests showed the choice of normalisation band did not significantly affect our final results.
The redshift, normalisation, and $\nu_{0}$ priors are broad and chosen to ensure it is unlikely that these parameters lie outside this range; the temperature prior was chosen to ensure the dust temperature is above the CMB temperature, and generally reflects the known distribution of dust temperatures in DSFGs \citep{Chapman2005, Casey2014, Clements2018}, and the $\beta$ prior is typical of what is found in the literature \citep{Bianchi2013, Casey2014}.
We ran experiments using different and more informative priors, but found that in general we were often reproducing our prior, justifying our choice of an uninformative prior.

\begin{figure*}
    \centering
    \includegraphics[width=1\linewidth]{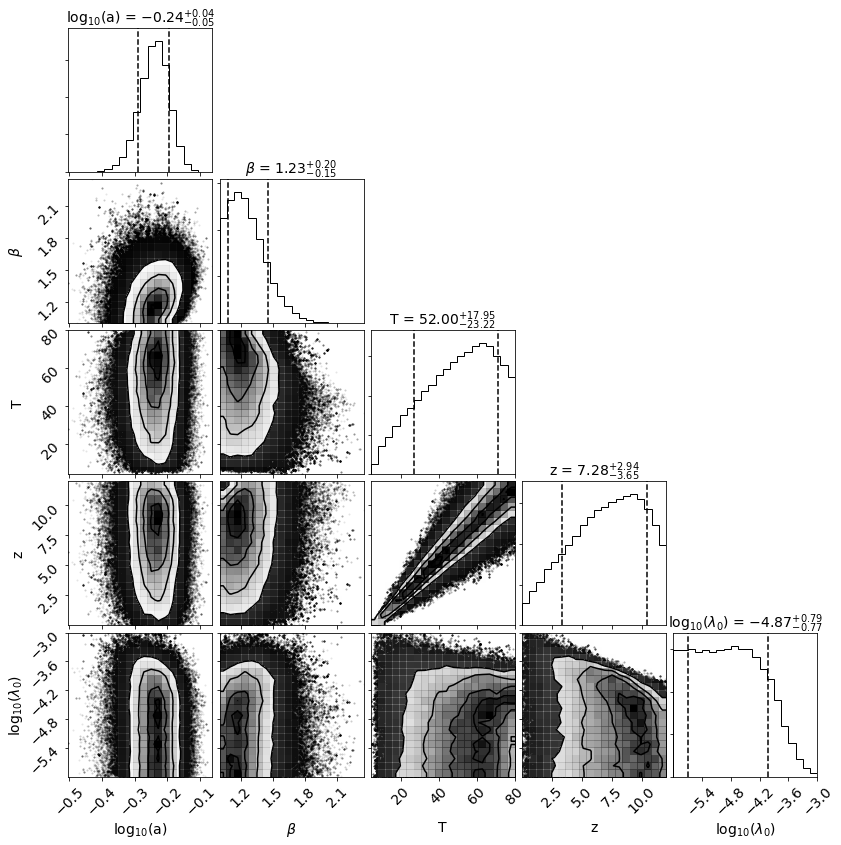}
    \caption{The samples and marginalised posteriors obtained after fitting the model described in Equation \ref{eq:MBB} to the photometry from NGP6\_D1. Median values are given above each parameter, whilst errors are taken from the 14th and 86th percentile of each marginalised posterior. The vertical dashed lines also show the 14th and 86th percentiles.}
    \label{fig:samplesNika}
\end{figure*}

\begin{figure}
    \centering
    \includegraphics[width=1\linewidth]{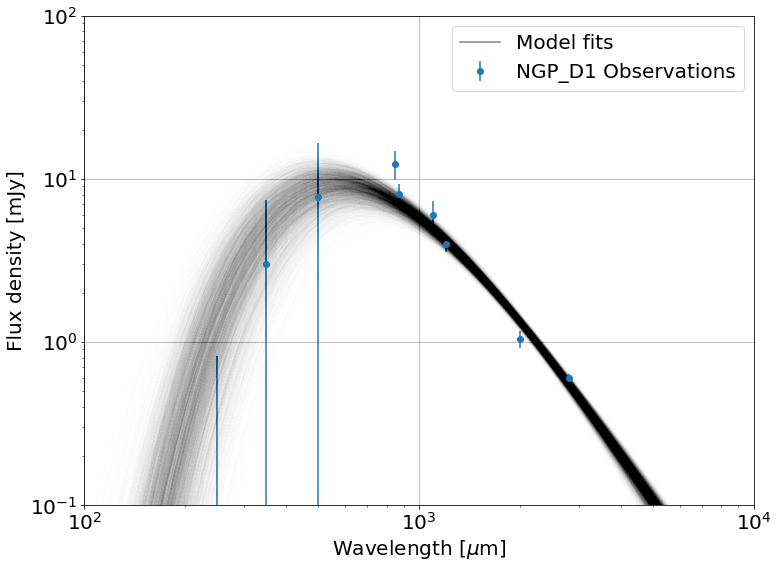}
    \caption{3000 randomly chosen model fits from Figure \ref{fig:samplesNika}, compared to observations of NGP6\_D1}
    \label{fig:samplesNika2}
\end{figure}

To perform our fit, we use 100 walkers over 10,000 steps, throwing away the first 1,000 samples in each chain as a burn-in phase and manually examining the chains to ensure that the samples have fully burnt-in.
Figure \ref{fig:samplesNika} shows the results of the fit, using the same photometry data in Table \ref{table:photometry}.
The temperature-redshift degeneracy can clearly be seen, and indicates that, as expected, we are unable to constrain either the redshift or temperature individually (though we are able to constrain their ratio reasonable well).
The $\nu_{0}$ parameter generally favours $c/\nu_0 < 100 \mu$m, indicating that our fits are well matched by an optically thin model.
Our normalisation suggests that the observed SCUBA-2 flux density of NGP6\_D1 is higher than its true value, in agreement with our SMA observations.
The $\beta$ values are lower than many other $z > 4$ sources in the literature \citep{Riechers2013, Fudamoto2017}, but within the expected range (though removing the NIKA data can raise this value, as is shown in Appendex \ref{apen:one}).
Additionally, in Figure \ref{fig:samplesNika2}, we plot 3000 single modified blackbody fits to the model, with parameters chosen at random from the samples in the posterior. 
As expected, most of the uncertainty lies in the SPIRE bands, where our constraints are weakest.

\subsection{FIR luminosity, SFR, and dust mass}

\begin{figure}
    \centering
    \includegraphics[width=1\linewidth]{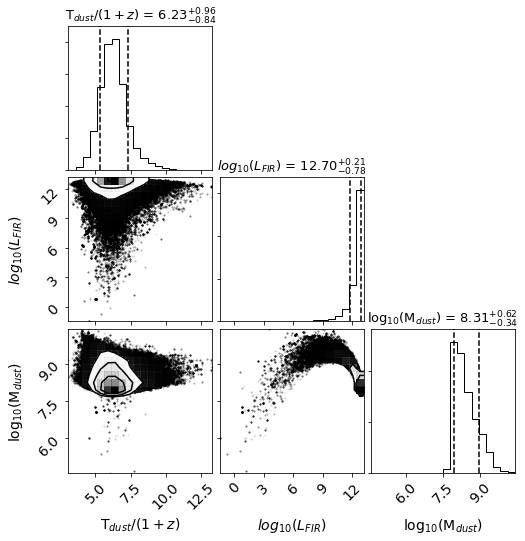}
    \caption{Derived parameters for NGP6\_D1, using the samples from Figure \ref{fig:samplesNika}}
    \label{fig:newsamplesNika}
\end{figure}

We calculate the FIR luminosity by integrating between 42.5 and 122.5 $\mu m$ on the resulting rest-frame FIR SED produced using the parameters from each of the 9,000 samples shown in Figure \ref{fig:samplesNika}.
We additionally calculate the dust mass, for which we follow \citet{Riechers2013} and use 
\begin{equation}
\label{eq:dmass}
    M_{dust} = S_{\nu} D_{L}^2[(1+z) \kappa_{\nu}B_{\nu}(T)]^{-1}\tau_{\nu}[1-exp(-\tau_{\nu})]^{-1},
\end{equation}
where S$_{\nu}$ gives the rest-frame flux density at 125 $\mu m$, D$_{L}$ is the luminosity distance, $\kappa_{\nu}$ is the mass absorption coefficient and is assumed to be $\kappa_{\nu} = 2.64$ m$^2$kg$^{-1}$ at 125 $\mu m$ \citep{Dunne2003}.
In Figure \ref{fig:newsamplesNika} we show our results.
This method was also tested on photometry from HFLS3 \citep{Riechers2013}, excluding the redshift, and we found that the literature values of these parameters were generally within the 14th - 86th percentiles of our predictions.

We next examine the fitted parameters and results derived from them, and compare our results for NGP6\_D1 to other objects in the literature.
The T$_{dust}$/($1+z$) parameter of NGP6\_D1 is lower than is generally seen in the 500 $\mu m$ risers in Table 3 of \citet{Dowell2014} and Figure 8 of \citet{Asboth2016}. These have typical values of around 9 - 12, with only one source, FLS 32 in \citet{Dowell2014}, having a comparable T$_{dust}$/($1+z$) = $6.7$ $\pm$ $3$.
However, our result is consistent with the $z > 4$ \citet{Ivison2016} selected sources, the current spectroscopically confirmed sources of which are listed in \citet{Fudamoto2017}, and have an average T$_{dust}$/($1+z$) parameter of 6.05 $\pm$  $0.44$, in much better agreement with our result for NGP6\_D1.
The spectroscopically confirmed \citet{Chapman2005} sources have temperatures fit using single temperature modified blackbodies, though with a fixed $\beta$ value of $\beta = 1.5$.
They find a mean T$_{dust}$/($1+z$) of 12.3 $\pm$ 3.0, once again significantly higher than we have found for NGP6\_D1, with no sources where T$_{dust}$/($1+z$) $< 8$. 
The redshift distribution of their sources is also limited to $z < 4$, with most of their sources at $2 < z < 3$.
The predicted FIR luminosity of NGP6\_D1 is reasonably well constrained, with $log_{10}(L_{FIR}) = 12.70^{+0.21}_{-0.78}$, where the errors give the 14th and 86th percentiles of the posterior distribution.
These values suggest NGP6\_D1 is likely a ULIRG, and if it is at $z > 4$ as our observations suggest, it is likely one of the least luminous detected $z > 4$ DSFGs to date (see Table 7 of \citet{Fudamoto2017} for a comparison of several literature $z > 4$ DSFGs and their derived properties). It may be more representative of the general $z > 4$ DSFG population.
We convert this FIR luminosity to a SFR by using Equation 4 of \citet{Kennicutt1998}, and convert to a Kroupa IMF by dividing by 1.5, as described in \citet{Schiminocich2007} (see also \citet{Hayward2014}).
This gives 
\begin{equation}
SFR [M_{\odot}~ yr^{-1}] = 1 \times 10^{-10} L_{FIR}~ [L_{\odot}],
\end{equation}
which leads to a predicted SFR for NGP6\_D1 of 512$^{+301}_{-426}$ $M_{\odot}~yr^{-1}$.
This value is an order of magnitude lower than almost all other non-lensed $z > 4$ DSFGs (see Table 7 of \citet{Fudamoto2017}), with the notable exception of HDF 850.1 \citep{Walter2012}, which has a SFR corrected for lensing (using the magnification estimated by \citet{Neri2014}) of $\sim 530$ $M_{\odot}~yr^{-1}$. 
We find this result notable because HDF 850.1 is also the only other SPIRE dropout in Table 7 of \citet{Fudamoto2017}, indicating that SPIRE-dropouts may represent the more populous lower luminosity DSFGs at $z \geq 4$.

The predicted dust mass for NGP6\_D1 is between $10^8$ and $10^9$ $M_{\odot}$, slightly lower than, but comparable to, other literature values (Table 6 of \citet{Fudamoto2017}). 
Of the 6 dust masses presented by \citet{Fudamoto2017}, only one (G09-83808c) is as low as the predicted value for NGP6\_D1, and this one source is additionally gravitationally lensed by a factor of 8.2 $\pm$ 0.3 \citep{Oteo2017a}.

\subsection{CO lines, CO luminosity and gas mass} 

\begin{figure}
    \centering
    \includegraphics[width=\linewidth]{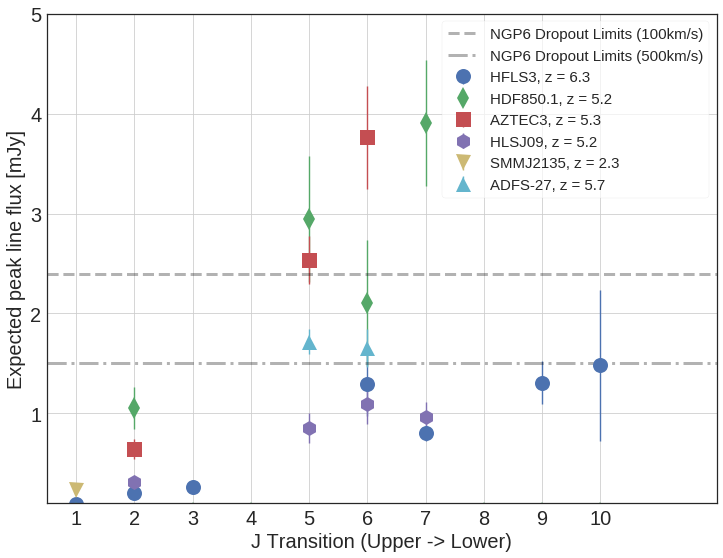}
    \caption[Expected CO line flux densities]{The expected peak line flux of the CO line transitions of NGP6\_D1, as estimated by several well studied high-z DSFGs and indicated using the different coloured markers. Each marker represents the peak line flux for that DSFG and at that J transition. The dashed line shows the upper limits at 100 $km~s^{-1}$, whilst the dot dashed line shows the upper limits at 500 $km~s^{-1}$}
    \label{fig:CoLims}
\end{figure}

Our spectroscopic observations from both EMIR and the RSR generally cover from 73 - 114 GHz to a similar RMS of around 0.5 - 0.7 mJy. We estimated the expected CO line flux densities for NGP6\_D1 by multiplying the observed 850 $\mu m$ flux density of NGP6\_D1 by the CO line flux to 850 $\mu m$ continuum ratio in several other high redshift DSFGs.
In Figure \ref{fig:CoLims}, we plot these estimates for six well studied DSFGs, as well as the detection limits of our EMIR and RSR observations.
As our detection limit is dependent on the assumed rotational velocity of NGP6\_D1 (which determines to what velocity we smooth our data), we plot an estimate for both 100 km/s and 500 km/s with the later being typical for high-z DSFGs \citep{Bothwell2013, Fudamoto2017}.

\begin{figure}
    \centering
    \includegraphics[width=\linewidth]{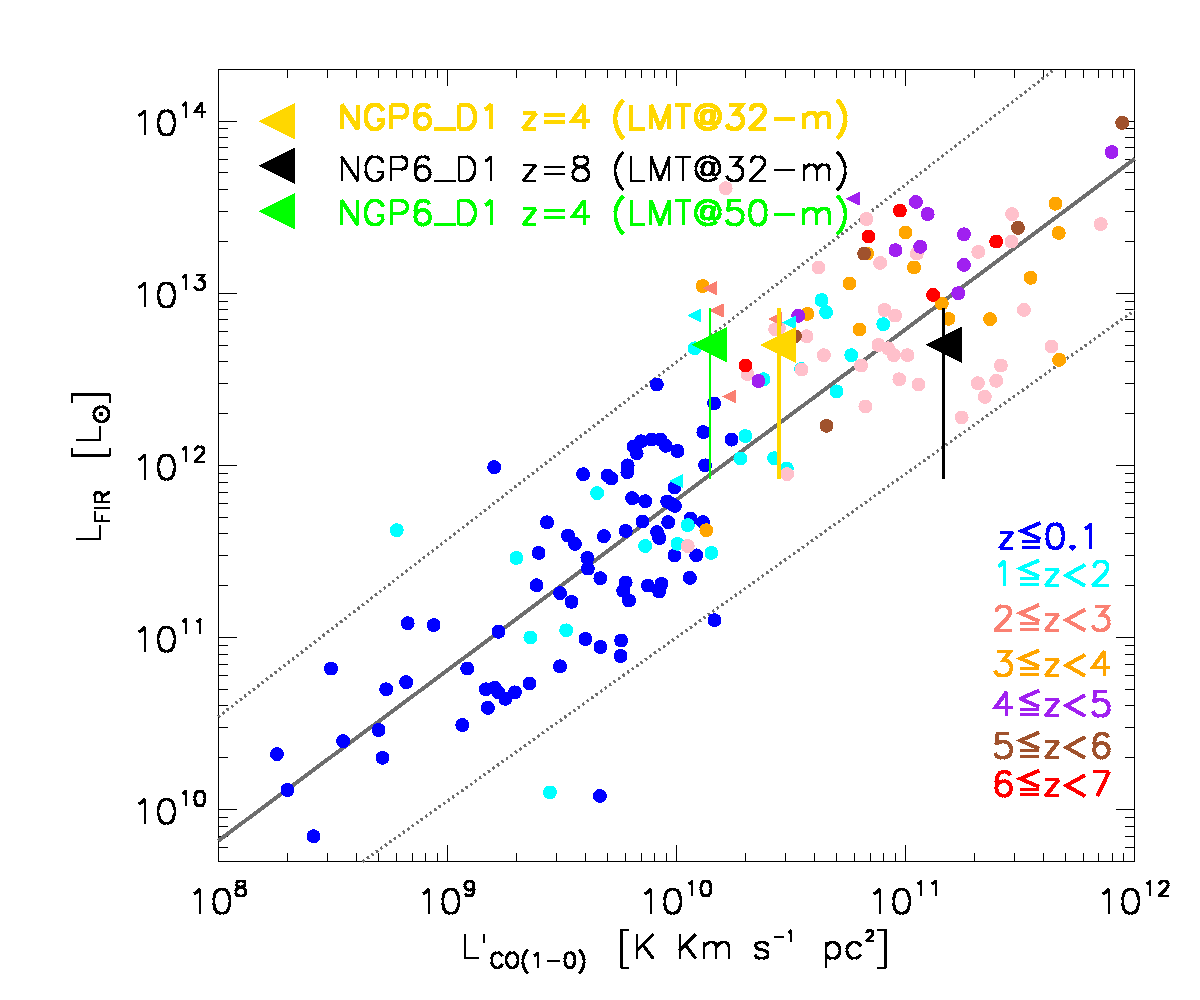}
   \caption{L$'_{\text{CO(1--0)}}$-- L$_{\text{FIR}}$ correlation for (U)LIRGs ($z\leq$0.1, \citealt{Papadopoulos2012}) and DSFGs ($z\geq$1, \citealt{Riechers2010, Harris2012, Bothwell2013, Gullberg2018, Yun2015, Strandet2017, Riechers2017, Yang2017, Jimenez-Andrade2018, Zavala2017, Aravena2016}) from the literature at different redshifts (colored circles and legends). Overplotted is the parameterized L$'_{\text{CO(1--0)}}$-- L$_{\text{FIR}}$ relation proposed by \citealt{Greve2014} (gray solid line) with the associated scatter of the data (gray dotted lines). We show our CO luminosity upper limits for $z$=4 and $z$=8 solution represented by the yellow and black triangles, respectively, with vertical error bars of the size of FIR luminosity uncertainty obtained in section 3.2. We also show the same $z$=4 solution for a non-detection but observed with RSR at LMT@50-m (green triangle) which will be near twice the depth of our current RSR observations at  LMT@32-m}
    \label{fig:co_upper}
\end{figure}

At the redshifts estimated from template fitting, between 72 - 114 $GHz$ we expect to see the CO(5-4) - CO(8-7) transitions.
These are typically the brightest CO lines seen in DSFGs (see Figure \ref{fig:CoLims}).
Given no detection, we place 3$\sigma$ upper limits of 2.4 $mJy$ at a resolution of 100 $km~s^{-1}$ and 1.5 $mJy$ at a resolution of 500 $km~s^{-1}$.
Figure \ref{fig:CoLims} suggests we can rule out a CO line flux in NGP6\_D1 that is similar to AZTEC3 \citep{Riechers2010} or HDF 850.1 \citep{Walter2012}.
We may be able to marginally rule out a line flux similar to ADFS-27 \citep{Riechers2017}, under the assumption that the line widths in NGP6\_D1 are $\sim$ 500 $km~s^{-1}$.
We cannot however rule out a spectral line energy distribution (SLED) similar to HLSJ09 \citep{Combes2012} or HFLS3 \citep{Riechers2013}.
Our observations therefore approach limits that suggest that NGP6\_D1 might be CO deficient compared to other high-z DSFGs.

We also place our CO luminosity upper limit on the observed L$'_{\text{CO(1--0)}}$-- L$_{\text{FIR}}$ correlation for galaxies with high star formation efficiencies. We follow Equations 6 and 7 from \citealt{Bothwell2013} to estimate CO luminosity upper limits using the RMS noise per 31 MHz channel of RSR spectrum, a mean linewidth of 500 km/s (typically expected for DSFGs) and adopting several redshift solutions from 0 to 10. Figure \ref{fig:co_upper} shows the L$'_{\text{CO(1--0)}}$-- L$_{\text{FIR}}$ correlation for (U)LIRGs ($z\leq$0.1) and DSFGs ($z\geq$1) from the literature at different redshifts.

In figure \ref{fig:co_upper} we plot our CO luminosity upper limit for a $z$=4 solution represented by the yellow triangle at the FIR luminosity obtained in section 3.2. For higher $z$ solutions, the upper limit moves towards the right (e.g. see black triangle for $z$=8 upper limit). It is worth noting that 4<$z$<8 solutions are well located within the scatter of the correlation of the ULIRG luminosity regime, as expected. 
Despite only a handfull of $z$>4 DSFGs being located within the ULIRG regime on the L$'_{\text{CO(1--0)}}$-- L$_{\text{FIR}}$ diagram, NGP6\_D1 upper limits suggest that this source could have similar properties to other $z$>4 DSFGs like ALESS65.1 ($z$=4.4, \citealt{Huynh2017}), AzTEC/C159 ($z$=4.6, \citealt{Jimenez-Andrade2018}), HDF850.1 ($z$=5.2, \citealt{Walter2012}), SDSSJ1044--0125 ($z$=5.8, \citealt{Wang2013}) and G0983808 ($z$=6.0, \citealt{Zavala2017}). 

If NGP6\_D1 lies at z$<$4, a CO detection would be expected, though the scatter in figure \ref{fig:co_upper} means that a non-detection in our current dataset remains a possibility. Deeper spectroscopy of this source, with LMT@50-m, for instance, would exclude this possibility.

Our upper limits at z$>$4 lead to a molecular gas mass upper limit for NGP6\_D1 of $\sim$1 $\times$ 10$^{11}$ M$_{\odot}$ and a  upper limit to the gas depletion time, $\tau_{dep} = $ M$_{H2}$/SFR of $\sim$800 Myr which includes the $\sim $ 100 Myr depletion times seen in DSFGs and barely reject $\sim$ 1 Gyr depletion times seen in normal\footnote{i.e. not mergers or quasars, which are more typically studied at $z$ > 1} $z$ > 1 galaxies \citep{Tacconi2010, Bothwell2013, Carilli2013}.


\section{Discussion}
\label{sec:discussion}

\subsection{Comparison to the Literature}
In this Section, we compare NGP6\_D1 to other dropout-like sources in the literature.
Only recently have large surveys at 850 $\mu m$ been completed, so few examples of 850 $\mu m$ risers or SPIRE dropouts have been published to date. 

\citealt{Ikarashi2017} identify and characterise two sources, selected partially on the basis of their faint SPIRE emission.
These sources both are undetected in SPIRE, but are both detected by SCUBA-2 at 850 $\mu m$ and ALMA at 1.1 $mm$, with flux densities of $\sim 4.5$ and $\sim 3.0~mJy$ each in the respective bands.
One source, ASXDF1100.053.1, is further detected by the VLA at 6 $GHz$, with a flux density of $4.5 \pm 1.1$ $ \mu~Jy$.
Compared to NGP6\_D1, these sources are 4$\times$ fainter at 850 $\mu m$, despite neither NGP6\_D1 or either of the \citealt{Ikarashi2017} sources being detected in SPIRE.
Furthermore, NGP6\_D1 is 4$\times$ brighter at 6 $GHz$ compared to ASXDF1100.053.1.

\citealt{Boone2013} found a SPIRE-dropout during APEX/Laboca follow up in the Herschel Lensing Survey \citep{Egami2010}.
They conclude that it is possibly a low luminosity source ( $L_{FIR} < 10^{12}~L_{\odot}$ ) at z $>$ 4 that is being lensed, possibly multiple times, by the brightest cluster galaxy in AS1063 (RXC J2248.7-4431). 
They further postulate this dropout source may be associated with an optically detected z $=$ 6.107 system.
Further follow up work by \citealt{Boone2016} reveals numerous dropout sources amongst the Herschel Lensing Survey fields, with ALMA and NOEMA programs underway to determine the nature of these sources.
The key difference between NGP6\_D1 and the dropouts found in the Herschel Lensing Surveys is that there is no evidence that NGP6\_D1 is being lensed by any structure. 

ADFS-27 is a dusty major merger and a 850 $\mu$m riser ( S$_{850}$ $>$ S$_{500}$ $>$ S$_{350}$) at $z=5.655$ \citep{Riechers2017}.
As ADFS-27 has its observed SED peak at $\sim$ 850 $\mu$m, a lower luminosity variant would likely still be detected by SCUBA-2, but remain undetected by SPIRE. A fainter version of ADFS-27 would thus be classed as a SPIRE dropout, similar to NGP6\_D1. 

\subsection{What are the SPIRE Dropouts?}

Given what we have learned about NGP6\_D1, we here examine other populations that may be similar.

The 850 $\mu m$ risers ($S_{250} < S_{350} < S_{500} < S_{850}$ - often just the last of these is used due to non-detection in the shorter wavelength SPIRE bands) may represent a population of DSFGs at redshifts $z > 6$  \citep{Ikarashi2017, Riechers2017}.
The idea behind this is similar to the 500 $\mu m$ riser population \citep{Dowell2014, Asboth2016, Ivison2016}; at $z \geq 6$ the rest-frame $\sim 100$ $\mu m$ peak of dust emission would be redshifted into the 850 $\mu m$ band.
A source bright enough to be detected at both 850 and 500 $\mu m$ would then be classed as an 850 $\mu m$ riser.
This population potentially relates to NGP6\_D1; a source with a 500 $\mu m$ flux density below the nominal SPIRE detection threshold, but still detected at 850 $\mu m$ would be classed as a SPIRE dropout.

Few confirmed 850 $\mu m$ risers are known.
As part of a follow up of 500 $\mu m$ risers, \citet{Riechers2017} discovered ADSF-27, a binary HLIRG 850 $\mu m$ riser.
It has a spectroscopically confirmed redshift of $z = 5.655$ and a luminosity of $2.4 \times 10^{13}$ $L_{\odot}$.
Despite this high luminosity, ADFS-27 is only just bright enough to be detected in the SPIRE bands in the deepest Herschel surveys \citep{Riechers2017}.
These authors suggest that the surface density of 850 $\mu m$ risers could be as low as $9 \times 10^{-3}$ $deg^{-2}$, if ADFS-27 remains the only 850 $\mu m$ riser amongst the SPIRE-only detected 500 $\mu m$ risers.
The rarity of 850 $\mu m$ risers is supported by \citet{Ivison2016}, who followed up a sample of 109 red SPIRE sources from the H-ATLAS survey with SCUBA-2, and found no 850 $\mu m$ risers.

A key difference between the \citet{Ivison2016} sample and ADFS-27 however is the flux density at 500 $\mu m$; whilst the \citet{Ivison2016} sample had a minimum 500 $\mu m$ flux density of 30 $mJy$ from completeness considerations, the 500 $\mu m$ flux of ADFS-27 is only $24.0 \pm 2.7~mJy$.
Indeed, HDF-850.1 \citep{Walter2012}, the only other well studied SPIRE dropout, is undetected in SPIRE, with a 500 $\mu m$ flux density $< 21~mJy$.
What luminosity would a typical DSFG have to have in order to be detected in SPIRE at ($S_{500} > 30~mJy$), and be an 850 $\mu m$ riser ($S_{850} > S_{500}$)?
In the top panel of Figure \ref{fig:pepsi} we plot the luminosity, redshift and dust temperature a source would need to be detected in both SPIRE at 500 $\mu m$ and SCUBA-2 at 850 $\mu m$, whilst also having $S_{850} > S_{500}$.
We would not expect to see many 850 $\mu m$ risers at $z < 4$, as they would require cold dust temperatures of $< 30$K. 
Using Equation \ref{eq:dmass} these requirements would lead to dust masses $> 10^{10} M_{\odot}$, 2 orders of magnitude higher than seen typically in the literature \citep{daCunha2015}.
At $z > 5$ however, we would also expect sources to be rare, as only the most luminous HLIRG and above systems with dust temperatures of 40 - 50K would be detected as 850 $\mu m$ risers.
These results seem to contrast with the observed $T_{dust}/(1+z)$ of ADFS-27, with $T_{dust}/(1+z) = 8.3$ at $z = 5.655$.
However it should be noted that ADFS-27 is a merger of two systems, with a separation of around 10 $kpc$. 
Even though they are at the same redshift, it is possible to construct a viable 850 $\mu m$ riser SED; experiments show that fitting dual single modified black bodies to the two components of ADFS-27, with $\sim$20 K and $\sim$50 K dust temperatures, can accurately reproduce the observed SED of the dual system.

The SPIRE dropouts may also be fainter analogues of the 500 $\mu m$ risers; a 500 $\mu m$ riser too faint to be detected in the SPIRE bands may still be detected at 850 $\mu m$ due to the different depths SPIRE and typical 850 $\mu m$ instruments reach.
Indeed, given the depths reached  in our observations, it is entirely plausible that NGP6\_D1 is merely a 500 $\mu m$ riser as opposed to an 850 $\mu m$ riser.
In the bottom panel of Figure \ref{fig:pepsi}, we plot a SPIRE dropout with an 850 $\mu m$ flux density of 10 $mJy$.
We further indicate where, in the plot of $T_{dust}/(1+z)$ vs z, such a source would be detected in SPIRE (and therefore not be a dropout), where it is a fainter version of a 500 $\mu m$ riser, and where it is a fainter version of an 850 $\mu m$ riser.
For a source with an 850 $\mu m$ flux density of 10 $mJy$, about half of the parameter space would not be detected in SPIRE, including ULIRGS with $z \gtrsim 4$ and/or T$_{dust} < 50$ K sources.

Comparing the two panels of Figure \ref{fig:pepsi}, it is immediately apparent that the SPIRE dropouts cover a much larger range of parameter space compared to the 850 $\mu m$ risers seen in the top panel.
Furthermore, this selection is better at sampling the lower luminosity population; it is able to select sub-HLIRG objects with dust temperatures of 30 - 50 K, as seen for example, in the population studied by \citet{Chapman2005} and \citet{Miettinen2017}.
The SPIRE detected 850 $\mu m$ risers on the other hand, are limited to HLIRG-like objects at $z > 5$, and below $z = 4$ are limited to cold $T_{dust} < 30$ K objects.
If the trends seen at $z = 2 - 3$ in \citet{Chapman2005} and \citet{Miettinen2017}, that \textit{most} SMGs have dust temperatures $\sim 30$ - 50 K, continues to $z > 4$, then the SPIRE dropouts could well represent a population of medium dust temperature (T$_{dust}$ = 30 - 50 K), ULIRG-like objects at $z > 4$. For all reasonable luminosity functions, these sources will be more numerous than the high luminosity HLIRGs.
This kind of source would be inaccessible in the optical/NIR without the benefit of negative k-correction, and be inaccessible to SPIRE because of the faint emission in the observed-frame FIR.

\subsection{The Nature of NGP6\_D1}

\begin{figure*}
    \centering
    \includegraphics[width=0.95\textwidth]{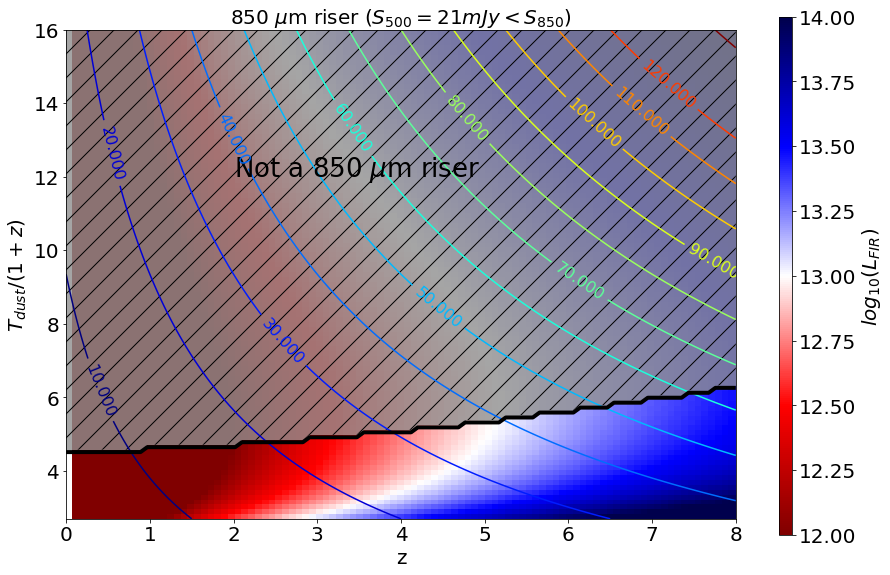}
    \includegraphics[width=0.95\textwidth]{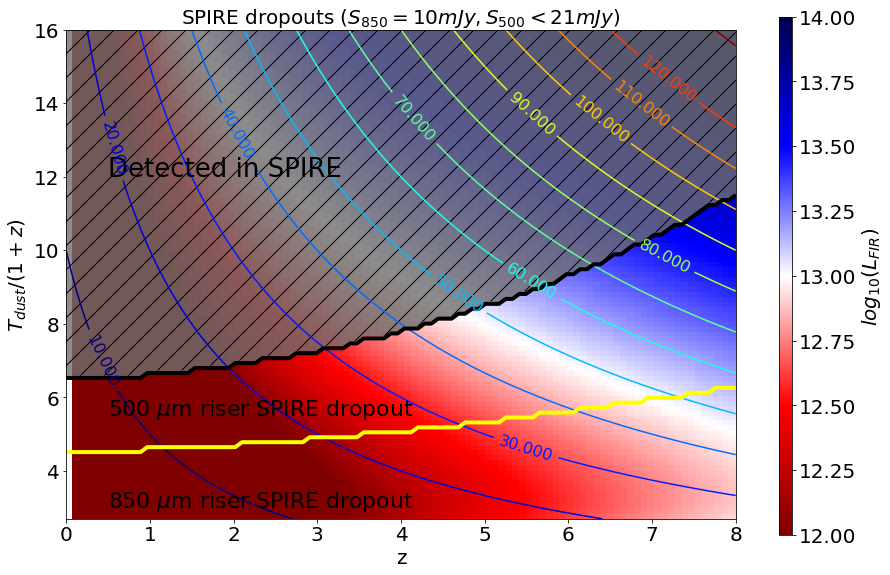}
    \caption[Luminosity limits of SPIRE dropouts and 500 $\mu m$ risers]{\textit{(Top)}: $T_{dust}/(1+z)$ against z for a single modified blackbody with $\beta~ =~ 2.$ and $\nu_{0}~ =~ 100 \mu$m, normalised to $S_{500}~ =~ 21~mJy$. The shaded region shows where this source is \textit{not} a 850 $\mu m$ riser, whilst the background colours show the luminosity of such a source as a function of z and $T_{dust}$. \textit{(Bottom)}: The same as the top, but for a SPIRE dropout with normalisation $S_{850} = 10~mJy$. The shaded region shows where the source is \textit{not} a SPIRE dropout, whilst the yellow line separates out SPIRE dropouts which are also 500 $\mu m$ risers from those which are 850 $\mu m$ risers. The lower y limit is constrained by the CMB temperature as $y = 2.73 \times (1+z)$, whilst the upper is chosen to broadly fit sources from the literature. }
    \label{fig:pepsi}
\end{figure*}

We now return to the central question of this paper, what kind of object is NGP6\_D1?
It is difficult to say with certainty; whilst labelling NGP6\_D1 and the SPIRE dropouts in general as a likely population of $z > 6$ DSFGs is attractive and a viable possibility, it is also possible that NGP6\_D1 and the SPIRE dropouts are examples of a cooler, 30 - 50K population of DSFGs that exist at $z = 3 - 6$.
It is unlikely that SPIRE alone can be of much help in accessing the $z > 5$ population of DSFGs, as Figure \ref{fig:pepsi} clearly demonstrates that it cannot detect many sub-HLIRG objects at $z > 5$ unless they are lensed.


We now examine the possible nature of NGP6\_D1, and attempt to rule out the least likely scenarios.

\textit{(1) Galactic}. 
We do not detect NGP6\_D1 in the optical/NIR down to AB magnitudes of 22 - 19.
Under the assumption that our source is at $z = 0$, the FIR SED constrains the dust temperature of our source to be $<$10K (see Figure \ref{fig:samplesNika}), and our observations from NOEMA and the SMA constrain the size of the source to be $< 1$ light year if NGP6\_D1 is within 30kpc of the Earth. 
NGP6\_D1 could therefore be a giant molecular cloud (GMC), but this is unlikely for the following reasons.
NGP6\_D1 was detected in the northern galactic pole, where we do not expect significant contamination from galactic sources or from the disk of the Milky Way. 
Its temperature would be comparable to, or lower than, the cores of GMCs \citep{Schneider2014}, and sources do exist with temperatures lower than the CMB, such as the Boomerang nebula \citep{Sahai1997}.
However, unless NGP6\_D1 is at a distance of 30kpc, its size is smaller than that of other molecular clouds, which are typically around 1 light year across \citep{Murray2010}. 
A system this cold and small would be very short lived. 
This, combined with the lack of any extended structure around NGP6\_D1, and the lack of a detection in WISE or IRAS indicate that a galactic origin is unlikely.

\textit{(2) Intermediate redshifts (z = 0 - 4)}.
In their examination of 73 850 $\mu$m selected sources, \citealt{Chapman2005} discover only 9 sources with $T_{dust} < 20$ K, all of which lie at z $<$ 1. 
\citealt{Cortese2014} also find many local ($< 30$ Mpc) sources with dust temperatures between 10 and 20 K, but no source with $T_{dust} < 10$ K.
If our source is a local ($z < 1$) galaxy, it would be one of the coldest galaxies in the Universe, with dust temperatures comparable to the Cosmic Microwave Background (CMB). 
Even between $z = 0 - 4$, the CMB varies in temperature between 2.7 and 13.5K.
Over the same redshift range, the temperature corresponding to the minimum $\chi^2$ for NGP6\_D1 varies between $\sim 2.5 - 22.5$K.
While the simple SED fits in Figure \ref{fig:samplesNika} indicate that a $z = 0 - 4$ solution is possible, consideration of the CMB temperature floor makes at least the lower half of this range highly implausible. The more physical template fitting method of Figure \ref{fig:SED} favours a high-redshift solution as do the existing results on similarly selected objects \citep{Riechers2013, Dowell2014, Ivison2016, Fudamoto2017}.
A z $\sim$ 2 solution thus seems unlikely.

Perhaps the most interesting possibility for an intermediate redshift solution is that NGP6\_D1 is similar to LSW 20, the 500 $\mu$m riser at a redshift of only $z = 3.3$ \citep{Dowell2014}. If such sources are common, they are not accounted for in existing template libraries but will still appear among red selected samples.
If this is the case, it would go some way to explaining discrepancies found when inferring general trends about the very red Herschel-SPIRE population, such as the over-abundance of red sources \citep{Dowell2014, Asboth2016, Bethermin2017}.


\textit{(3) High redshift (z = 4 - 8)}.
A $ 4 < z < 8$ solution would result in a dust temperature between 20 and 60 K, comparable to other high-z DSFGs.
Template fits from other well studied sources favour this solution, generally prefering the $z > 6$ solutions over $z < 6$.
The CO J(5-4), J(6-5), J(7-6), and J(8-7) should be visible in our spectrum, but as Figure \ref{fig:CoLims} shows, our RMS is not low enough that we can guarantee we should detect a line if our source is similar to HFLS3 or HLSJ09.
We thus conclude that a high redshift solution is the most likely explanation of NGP6\_D1.
Given the higher than expected radio flux (see Figure \ref{fig:SED}), and the fact that we do not detect any CO lines, we further suggest that NGP6\_D1 hosts an AGN, probably dust enshrouded, which contributes to the radio flux. 

\subsection{The SPIRE Dropout Population}


Figure \ref{fig:pepsi} suggests that SPIRE dropouts can inhabit a much larger range of luminosity-redshift-temperature parameter space than 850 $\mu m$ risers; the polygon that forms from the constraints that $20$ K $< T_{dust} < $ 80 K, and the approximate ``knee'' of the $z > 2$ DSFG luminosity function\footnote{ No 850 $\mu m$ riser nor SPIRE dropout has a dust temperature above 20 K below this redshift} at around $10^{13}$ $L_{\odot}$ \citep{Casey2014, Gruppioni2017, Koprowski2017} encompasses a much larger area of parameter space for the SPIRE dropouts compared to the 850 $\mu m$ risers. We therefore examine two of the largest extragalactic surveys at 850 $\mu m$ with significant Herschel-SPIRE survey overlap, to determine the number of SPIRE dropouts per $deg^2$.

\begin{figure*}
    \centering
    \includegraphics[width=\linewidth]{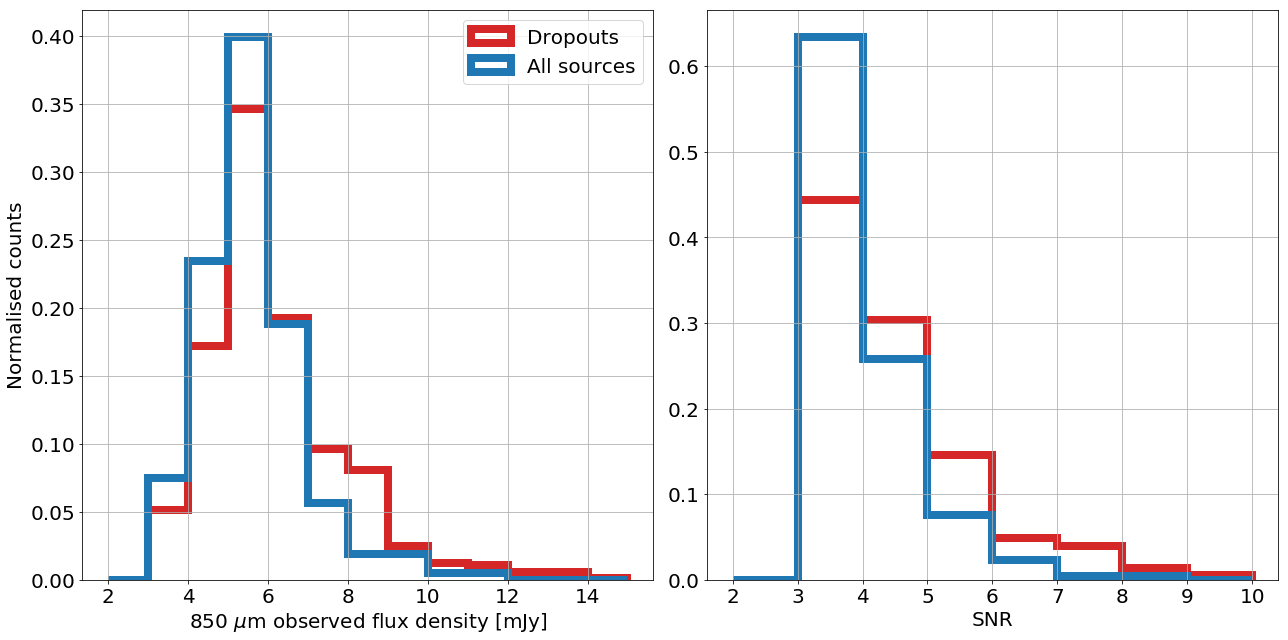}
    \caption[SPIRE dropouts number counts]{\textit{Left)} The number counts of both all detected SCUBA-2 sources in S2COSMOS (blue solid line), and dropouts (red dashed line showing sources uncorrected for boosting, red solid line for sources corrected for boosting. Bins have widths of 1 $mJy$. \textit{Right)} The SNR for the detections of both all detected S2COSMOS sources and dropouts.} 
    \label{fig:dropouts}
\end{figure*}

Initially, we searched for dropouts among the maps and catalogues from the observed $\sim$2 $deg^2$ COSMOS field of the SCUBA-2 Cosmology Legacy Survey S2CLS \citep{Geach2017}, using their first data release, which reached a uniform $1\sigma$ rms error of 1.6 $mJy$ across 2 $deg^2$ in the COSMOS field.
The S2COSMOS catalogues require a 3.5$\sigma$ detection for a source to be included in the catalogues, with a typical 1$\sigma$ value of 1.09 $\pm$ 0.24 $mJy$ at 850 $\mu m$.
We then matched these catalogues to catalogues from Herschel to search for any dropouts among the 719 detected SCUBA-2 objects.
For the Herschel catalogues, we used the HerMES \citep{Oliver2012} DR2 single band catalogues, where fluxes are extracted by the HerMES XID code \citep{Roseboom2010, Hurley2016} at positions found by the StarFinder code \citep{Diolaiti2000} at the corresponding wavelength.
We further assume Gaussian shaped beam FWHMs of 18.15, 25.15 and 36.3 arcsec at 250, 350 and 500 $\mu m$ respectively.
No attempt is made at cross-matching between bands, and three separate catalogues are made for the three SPIRE bands individually.
Using a search radius of 13.0 arcsec, equivalent to the beamsize of SCUBA-2, we cross-match the S2COSMOS sources with each of the three Herschel-SPIRE catalogues.
We find 213 sources which have no Herschel match in any of the three bands, a dropout fraction of 21.8\%.
If we use the beamsize from the Herschel 500 $\mu m$ band of 35.2 arcsec, we still find 57 dropouts (7.9\%).
Regardless of the precise beamsize, we find that a significant number of SCUBA-2 sources are dropouts.
In Figure \ref{fig:dropouts} we examine both the normalised and deboosted flux density distribution and normalised SNR distribution of the dropout sources when using the 13 arcsec search radius and compare this to the general SCUBA-2 population.
We find that the flux distributions of the dropouts and of the general population are broadly similar, with median values of 5.6 and 5.8 $mJy$ respectively, standard deviations of 1.3 and 1.8 $mJy$, and a long tail stretching towards higher flux densities.
This suggests that the dropouts are not merely the faint population of 850 $\mu m$ detected sources, but are a unique population of SMGs that remain undetected by Herschel.
Examining the SNR distribution, we find that 63\% of dropouts have a low SNR (with detection SNR $< 4$), compared to the general population, which has 44\% in this range.
This may imply that a number of the dropouts are noise spikes, but 11\% have a SNR $>$ 5 and, as we have shown here, at least some of the dropout population consists of real sources (26\% of all the S2CLS sources have a detection SNR $> 5$).
We detect ten dropouts with flux densities at 850 $\mu m$ $>$ 8 $mJy$ over $\sim$ 2 $deg^2$, corresponding to a source density of 5 $\pm$ 1.58 sources $deg^{-2}$, comparable to the 3.3 $\pm$ 0.8 bright red (S500 $>$ 30$mJy$) 500 $\mu m$ riser sources found by \citet{Dowell2014}.

These results are confirmed by Aguilar et al, (in preparation)
who make a comparison between AzTEC detected sources (i.e. S/N $>$ 3.5 at 1.1 $mm$) that were selected as 500 $\mu m$ risers or SPIRE dropouts on three well observed blank fields: GOODS-S, GOODS-N and COSMOS (270 sources on $\sim$0.86 $deg^2$ in total). 
They found that ~20\% of AzTEC sources were 500 $\mu m$ risers while $\sim$30\% were classed as SPIRE dropouts, similar to our results in S2CLS. 
After identification through radio-IRAC-CANDELS counterpart analysis and sub-mm SED fitting, they suggest that more than 50\% of this population is at z$>$4.
These results are in excellent agreement with our examination of S2COSMOS, and the predictions in Figure \ref{fig:pepsi}. The large numbers suggest these surveys are  detecting more ``normal'' DSFGs at $z > 4$, compared to the extreme sources detected by SPIRE \citep{Riechers2013, Dowell2014, Strandet2017, Zavala2017, Riechers2017}

\section{Conclusions}

\label{sec:conclusion}

NGP6\_D1 is a serendipitously detected SPIRE dropout, strongly detected at $\lambda \geq$ 850 $\mu m$ but not detected in any shorter wavelength bands.
Interferometric observations confirm it to be a single source, with no evidence for any optical or NIR emission, or nearby potentially lensing sources.
No $>3\sigma$ detected lines are seen in the spectrum of NGP6\_D1 across 32 $GHz$ of bandwidth, and the redshift remains unknown.
CO luminosity limits were calculated and these are consistent with the $L_{CO} - L_{IR}$ correlation of other $z>4$ DSFGs.
The degeneracy between the temperature and redshift of NGP6\_D1 prevents us from constraining either of these parameters strongly, but the luminosity and dust mass are reasonably well constrained, and suggest NGP6\_D1 is a ULIRG like object, with a dust mass $\sim 10^8$ - $10^9$ $M_{\odot}$ and a SFR of $\sim$ 500 $M_{\odot}~yr^{-1}$.
Template fitting over a range of galaxy types suggests the redshift of NGP6\_D1 is most likely between $z = 5.8$ and 8.3.
The upper limit on the gas mass of NGP6\_D1 suggests a maximum of $M_{H2}$ $ < (1.1~\pm~3.5) \times 10^{11}$ $M_{\odot}$, consistent with a gas-to-dust ratio of $\sim$ 100 - 1000.


We also find that SPIRE dropouts account for $\sim$ 20\% of all SCUBA-2 detected sources, but have similar flux density distributions to the general population.
We find that such dropouts likely represent either ULIRG like objects at $z > 4$, with dust temperatures around 30 - 50 K, comparable to those seen at $z = 2 - 3$, \textit{or} a population of $z > 6$ sources that have so far remained inaccessible to SPIRE.
These results are consistent with HDF 850.1 \citep{Walter2012}, one of the few well studied SPIRE dropouts, as well as the SPIRE dropouts identified by \citet{Ikarashi2017}, though the latter sources lack spectroscopic redshifts.

\section*{Acknowledgements}

The authors wish to recognize and acknowledge the very significant cultural role and reverence that the summit of Mauna Kea has always had within the indigenous Hawaiian community.  We are most fortunate to have the opportunity to conduct observations from this mountain.
The Herschel-ATLAS is a project with Herschel, which is an ESA space observatory with science instruments provided by European-led Principal Investigator consortia and with important participation from NASA. The H-ATLAS website is \verb!http://www.h-atlas.org/!.
The Submillimeter Array is a joint project between the Smithsonian Astrophysical Observatory and the Academia Sinica Institute of Astronomy and Astrophysics and is funded by the Smithsonian Institution and the Academia Sinica.
This work is based on observations carried out under project number 199-15, E16AD, and 227-14 with the IRAM NOEMA Interferometer [30m telescope]. IRAM is supported by INSU/CNRS (France), MPG (Germany) and IGN (Spain)
This research made use of Astropy, a community-developed core Python package for Astronomy \citep{Astropy2013}.
This work made extensive use of the Starlink Table/VOTable Processing Software, TOPCAT \citep{Taylor2005}.
This publication makes use of data products from the Wide-field Infrared Survey Explorer, which is a joint project of the University of California, Los Angeles, and the Jet Propulsion Laboratory/California Institute of Technology, funded by the National Aeronautics and Space Administration.
This research has made use of NASA's Astrophysics Data System Bibliographic Services.
This work made use of GILDAS \verb!http://www.iram.fr/IRAMFR/GILDAS!, a collection of state-of-the-art software oriented toward (sub-)millimeter radio-astronomical applications (either single-dish or interferometer).
DLC and JG acknowledge support from STFC, in part through grant numbers ST/N000838/1 and ST/K001051/1.
H.D. acknowledges financial support from the Spanish Ministry of Economy and Competitiveness (MINECO) under the 2014 Ram{\'o}n y Cajal program MINECO RYC-2014-15686.
This work would not have been possible without the long-term financial support from the Mexican Science and Technology Funding Agency, CONACYT during the construction and early operational phase of the Large Millimeter Telescope Alfonso Serrano, as well as support from the the US National Science Foundation (NSF) via the University Radio Observatory program, INAOE and the University of Massachusetts, Amherst.
LD, IO acknowledge support from the European Research Council Advanced Grant, COSMICISM. LD also acknowledges support from ERC consolidator grant, CosmicDust.
CY was supported by an ESO Fellowship.
MJM acknowledges the support of the National Science Centre, Poland through the POLONEZ grant 2015/19/P/ST9/04010 and SONATA BIS grant 2018/30/E/ST9/00208; this project has received funding from the European Union's Horizon 2020 research and innovation programme under the Marie Sk{\l}odowska-Curie grant agreement No. 665778.





\bibliographystyle{mnras}
\bibliography{library} 



\appendix

\section{The NIKA observations}
\label{apen:one}

During the course of our analysis, and as can be seen directly in Figure \ref{fig:samplesNika2} it became apparent that the NIKA fluxes appeared systematically lower than expected by a factor of $\sim$ 1.5.
Examination of the processed data, $\tau_{225~GHz}$ values during the observations, and observing logs do not suggest issues or any likely origin for any possible systematic errors.
The data were taken during a shared-risk mode, and the pipelines to reduce the raw data are no longer available, so it is not possible to re-reduce the data.
Nevertheless, comparison to data taken at other wavelengths appear to indicate a systematic offset beyond the reported errors, of around 50\%.

\begin{figure*}
    \centering
    \includegraphics[width=1\linewidth]{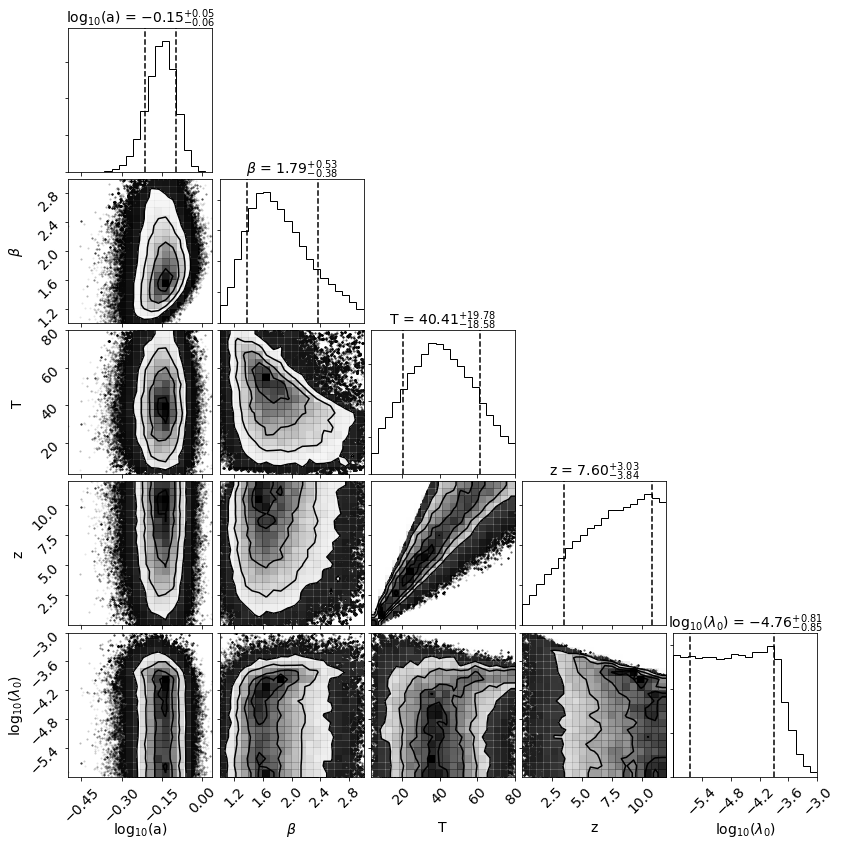}
    \caption{The same as Figure \ref{fig:samplesNika}, but excluding the NIKA data}
    \label{fig:sampleswithoutNika}
\end{figure*}

\begin{figure}
    \centering
    \includegraphics[width=1\linewidth]{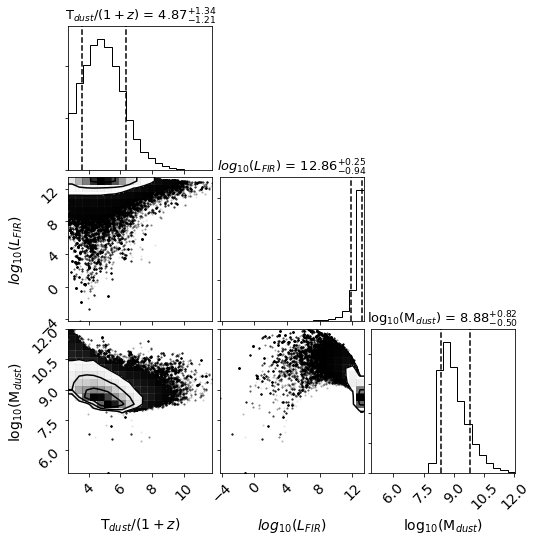}
    \caption{The same as Figure \ref{fig:newsamplesNika}, but excluding the NIKA data}
    \label{fig:newsampleswithoutNika}
\end{figure}

Because of these discrepancies, we also ran our sampler without the NIKA data included to see what effects it has on our results.
The results of excluding the data are shown in Figures \ref{fig:sampleswithoutNika} and \ref{fig:newsampleswithoutNika}.
The derived parameters end up similar, though parameters are slightly higher when excluding the NIKA data.
Both models are consistent with an optically thin model (i.e for all observed frequencies observations $\frac{\nu}{\nu_{0}} << 1$), have similar derived FIR luminosities ($\log_{10}(L_{FIR}/L_{\odot})$ = $12.86^{+0.25}_{-0.94}$ when excluding the NIKA data), and similar predicted dust masses ($\log_{10}(M_{dust}/M_{\odot})$ = $8.88^{+0.82}_{-0.50}$  when excluding).
The only clear differences are in the derived $\beta$ values, which are $\beta = 1.23^{+0.20}_{-0.15}$ when including the NIKA data but $\beta = 1.79^{+0.53}_{-0.38}$ when excluding the NIKA data, and in the T$_{dust}$/$(1+z)$ parameters, which when including the NIKA data are T$_{dust}$/$(1+z) = 6.22=3^{+0.96}_{-0.84}$ compared to T$_{dust}$/$(1+z) = 4.87^{+1.34}_{-1.21}$ without. 
Additionally, the reduced $\chi^2$ values for the median sampled parameters is $\chi^2_{red} = 3.95$ when including the NIKA data, but 2.62 when excluding it, indicating marginally better fits. 
This difference however does not appear to be having a significant effect on most of the derived parameters for NGP6\_D1, with primary differences emerging at the shortest ($\lambda < 500 \mu m$) wavelengths, where more data are required in order to resolve this potential conflict.
In this paper, we continue to include the NIKA data, but we note that it is possible that $\beta$ values may be higher, whilst T$_{dust}$/$(1+z)$ values might be lower.

\bsp	
\label{lastpage}
\end{document}